\documentclass[11pt]{article}
\usepackage[font=small,labelfont=bf]{caption}
\usepackage{listings}
\usepackage{color}
\usepackage{subcaption}
\usepackage{amsmath,amssymb}
\usepackage[left=1in, right=1in, top=1in, bottom=1in]{geometry}
\usepackage{verbatim}
\usepackage{titling}
\usepackage{parskip}
\usepackage{todonotes} 
\usepackage[british]{babel}
\usepackage[utf8]{inputenc}
\usepackage{slashed}
\usepackage{amsfonts}
\usepackage{amsthm}
\usepackage{amsmath,amssymb}
\usepackage{empheq}
\usepackage{xcolor}
\usepackage{hyperref}
\usepackage{nag}
\usepackage[utf8]{inputenc}
\usepackage{todonotes}
\usepackage{amsmath}
\usepackage{fullpage}
\usepackage{graphicx}
\usepackage{rotating}
\usepackage{float}

\usepackage{todonotes}
\usepackage{soul}

\title{The impact of network properties and mixing on control measures and disease-induced herd immunity in epidemic models: a mean-field model perspective}

\author{Francesco Di Lauro$^\dagger$, Luc Berthouze$^\triangledown$, Matthew D. Dorey$^\lozenge$, Joel C. Miller$^\ddagger$, and  Istv\'an Z. Kiss$^\dagger$\thanks{i.z.kiss@sussex.ac.uk}\\
$^\dagger$ Department of Mathematics, School of Mathematical and Physical Sciences,\\ University of Sussex, Falmer, Brighton BN1 9QH, UK\\[6pt]
$^\triangledown$ Department of Informatics, School of Engineering and Informatics\\ University of Sussex, Falmer, Brighton BN1 9QH, UK\\[6pt]
$^\ddagger$ Department of Mathematics and Statistics, School of Engineering and Mathematical Sciences\\ La Trobe University, Bundoora, Australia\\[6pt]
$^\lozenge$ Public Health and Social Research Unit, West Sussex County Council,\\Tower Street, Chichester, P019 1RQ, UK}
\date{May 2020}

\begin{document}
\maketitle

\begin{abstract}
The contact structure of a population plays an important role in transmission of infection.  Many ``structured models'' capture aspects of the contact structure through an underlying network or a mixing matrix.
An important observation in unstructured models of a disease that confers immunity, is that once a fraction $1-1/\mathcal{R}_0$ has been infected, the residual susceptible population can no longer sustain an epidemic.  A recent observation of some structured models is that this threshold can be crossed with a smaller fraction of infected individuals, because the disease acts like a targeted vaccine, preferentially immunizing higher-risk individuals who play a greater role in transmission.  Therefore, a limited ``first wave'' may leave behind a residual population that cannot support a second wave once interventions are lifted.
In this paper, we systematically analyse a number of well-known mean-field models for networks and other structured populations to shed further light on some important questions relevant to the current COVID-19 pandemic.  In particular, we consider the question of herd-immunity under several scenarios. When modelling interventions as changes in transmission rates, we confirm that in networks with significant degree heterogeneity, the first wave of the epidemic confers herd-immunity with significantly fewer infections than equivalent models with less or no degree heterogeneity. However, if modelling the intervention as a change in the contact network, then this effect might become much more subtle. Indeed, modifying the structure disproportionately can shield highly connected nodes from becoming infected during the first wave and therefore make the second wave more substantial. We strengthen this finding by using an age-structured compartmental model parameterised with real data and comparing lockdown periods implemented either as a global scaling of the mixing matrix or age-specific structural changes. Overall, we find that results regarding (disease-induced) herd immunity levels are strongly dependent on the model, the duration of the lockdown and how the lockdown is implemented in the model.

\end{abstract}
\section{Introduction}

The recent emergence of SARS-CoV-2 and the associated disease COVID-19 has had worldwide impact.  Many cities have had large outbreaks and brought them under control through major interventions.  Once those interventions are lifted, in absence of effective vaccination, a homogeneous model of infection spread would predict that as long as less than $1-1/\mathcal{R}_0$  of the population was infected, there is always a threat of resurgence. 

Despite large epidemics, cities such as New York remain well below the threshold expected to be be required to achieve herd immunity~\cite{stadlbauer2020seroconversion}.

To avoid the significant economic and health costs associated with continued interventions, it is natural to consider the so-called ``herd immunity'' strategy.  This strategy allows infection to spread with restrictions in place so that the outbreak finishes and interventions are lifted when the herd immunity threshold is reached.  Typically in an uncontrolled epidemic the herd immunity threshold is reached at the peak of the epidemic, and many additional infections occur as the outbreak slowly wanes.  The additional infections are sometimes referred to as the \emph{overshoot}~\cite{Handel2007}.  By calibrating the intervention so that there are no (or almost no) infections when the  herd immunity threshold is reached, interventions can be removed with minimal overshoot~\cite{di2020timing,morris2020optimal}. 

The severity of the epidemic in many places whose seroprevalence is still very low has led many to suggest that the herd immunity strategy is not tenable.  However, recent papers~\cite{Britton2020,gomes2020individual} suggest that immunity acquired through infection may be distributed through the population in such a way as to achieve herd immunity at a lower fraction affected than a homogeneous model would predict.  This is because the initial wave of infections preferentially affects those most at risk.  Thus it acts like a targeted vaccination, removing the people who are most likely to transmit infection from the susceptible pool. Generally, the time course of a real epidemic involves the following stages: (i) short period of unconstrained transmission, (ii) significant control or lockdown, and (iii) relaxation of control measures. Typically, during lockdown some spread persists and one pertinent question is whether relaxing the lockdown will lead to a second wave.  In~\cite{Britton2020}, this question was explored by looking at prevalence at the end of lockdown in scenarios where lifting the lockdown did not lead to resurgence. For the purposes of this paper, we refer to this fraction as the  disease-induced herd immunity (DIHI).

These papers make some simplifying assumptions about population structure that may not hold.  In particular they do not consider the fact that existing interventions tend to affect some contacts more than others.  For example, the transmission rates of household contacts are not significantly reduced (and in fact may increase) during interventions focused on reducing movement. Moreover, many of the highest risk positions in the disease network are in fact roles (health care workers, delivery drivers, teachers, etc) that would need to be maintained in most forms of a lockdown, meaning that some community links may be increased even as others are limited.

In this paper we will use several network-based models to improve our understanding of how population structure may affect the DIHI threshold.  Modelling epidemics on networks and the analysis of the resulting systems, be they stochastic or deterministic, is a well studied research area~\cite{Kiss2017,pastor2015epidemic,Porter2016}. The main reason for using networks is twofold. First, networks are intuitive and easy to understand and use, and second, they allow us to represent contact structure within a population to a high degree of accuracy. One of the most frequently studied questions in this area is how to understand and quantify the impact of various network properties on the invasion, spread and control of infectious diseases. Epidemics are usually modelled as stochastic process unfolding on the network. For example, the susceptible-infected-recovered/removed (SIR) epidemic on a network is modelled by the two separate processes:  infection and recovery. In the simplest case both are Poisson processes such that an infectious node $u$ connected to a susceptible node $v$ transmits at per-contact rate  $\tau$, causing $v$ to convert to infected.  Nodes recover  at rate $\gamma$ to an immune state independently of the network.

On a network with $N$ nodes, the SIR model leads to a continuous-time Markov Chain over $3^N$ states. Even for small values of $N$ such a system becomes tedious to handle (numerically or otherwise). This has led to a myriad of approximations in the form of mean-field models which focus on some average quantities to reduce the dimensionality of the resulting system~\cite{Kiss2017}. 

The mathematical analysis of epidemic models on networks (from regular and Erd\H{o}s-R\'enyi to scale-free networks such as the Barab\'asi-Albert preferential attachment model) has led to many elegant analytical~/~explicit results regarding the impact of network properties (e.g., degree heterogeneity, assortativity and clustering) on the epidemic threshold, final epidemic size or endemic equilibrium, optimal vaccination strategies~\cite{Porter2016,pastor2015epidemic,Chen2014,Holme2017}, as well as the fact that certain mean-field SIR models are exact in the limit of large configuration model networks~\cite{miller:ebcm_hierarchy,miller:equivalence}. Many such models are mathematically tractable and help us gain important intuition about how an epidemic spreads and what properties of the contact network affect this. While they may not directly inform policy making, they remain useful tools to develop intuition, highlight important population features and test various scenarios quickly and effectively.

In this paper we use four different mean-field models~\cite{Kiss2017} to approximate exact epidemics on networks: (a) degree-based heterogeneous mean-field, (b) clustered and unclustered pairwise, (c) a new edge-based-compartmental model that allows us to distinguish between household/local contacts and community/global contacts, and (d) an age-structured compartmental model parameterised with realistic age structure and contact matrices. 
The primary aim is to investigate how heterogeneity in model structure impacts DIHI. More importantly perhaps, we challenge the way lockdown has been implemented in many models, namely, by a simple reduction in $\mathcal{R}_0$ or the transmission rate while keeping the contact network or mixing matrix the same. We build a new edge-based-compartmental model able to distinguish between household and community  transmissions and use this to implement lockdown by either intervening on both types of connections or only on the community-based ones. In the same spirit, we use an age-structured compartmental model in which we implement lockdown either as a simple scaling of the entire mixing matrix or, more realistically, a set of age-specific structural changes.

The paper is structured as follows. In section~\ref{sec:methods} we describe all models including the underlying network types and relevant model and epidemic parameters. Section~\ref{sec:results} contains the results for the network models, whilst section~\ref{sec:results_SEIRD} provides the results for the age-structured model. Finally, in section~\ref{sec:discussion} we discuss the implications of our findings. Additional technical details are given in the Appendix.


\section{Methods}\label{sec:methods}
We consider a set of mean-field models. The first two are approximations of the exact stochastic SIR model on networks with heterogeneous degree distribution, without and with clustering. This is followed by a new edge-based-compartmental model with household structure, i.e., with the ability to distinguish between household and community contacts. Finally, we consider an age-structured SEIRD model based on realistic age-structure and mixing matrices. For the network-based mean-field models we choose a flexible degree distribution with good control of the mean and variance, i.e., the negative binomial. We loosely order the models by their relative complexity, corresponding to gradually incorporating more features of the underlying network or population contact structure. While the last model is not explicitly network-based, it uses realistic mixing matrices over an age-structured model based on UK data. One can consider this an extrapolation from (or an approximation of) an explicit network model where individual-level interactions are averaged out over groups of interest from an epidemiological view point. 

\subsection{Contact structure and epidemic model}
We consider a SIR/SEIRD epidemic spreading in a closed population of size $N=6.65\cdot 10^6$ (loosely the population size of the UK) with a well-defined contact structure. For illustrative purposes, we assume that the probability of an individual having $k$ contacts follows a negative binomial distribution
\begin{equation}
    P_{n,p}(k) = {k+n-1 \choose n-1} p^n (1-p)^k.
    \label{eq:neg-bin}
\end{equation}
The reason for this choice is that we want to highlight how heterogeneities in the contact structure play a central role in determining the DIHI. To illustrate this point, we consider three different scenarios for the degree distribution of the population. In all cases, we fix  $\langle k \rangle=n(1-p)/p$ and we use the remaining free parameter to tune the variance. 
To avoid individuals with degree $0$, the degree distribution is shifted by a constant $m$, thus making the effective average degree $\langle k \rangle = m + n(1-p)/p$. For normal-like and scale-free-like distributions we take $m=1$, and for the delta-like distribution we take $m=9$ (see Table~\ref{table:degrees}).

\begin{figure}
\centering
\includegraphics[scale=0.95]{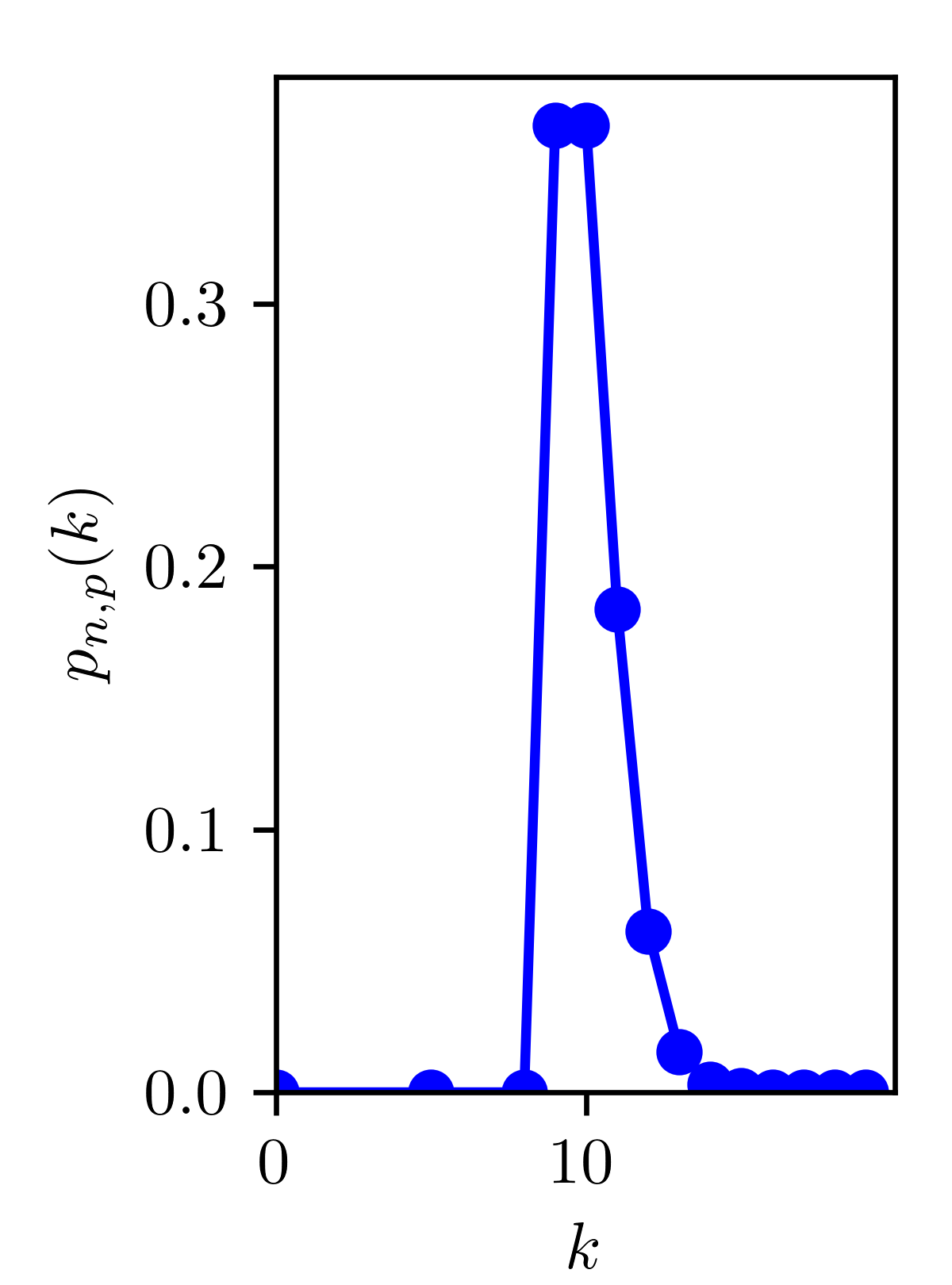}
\includegraphics[scale=0.95]{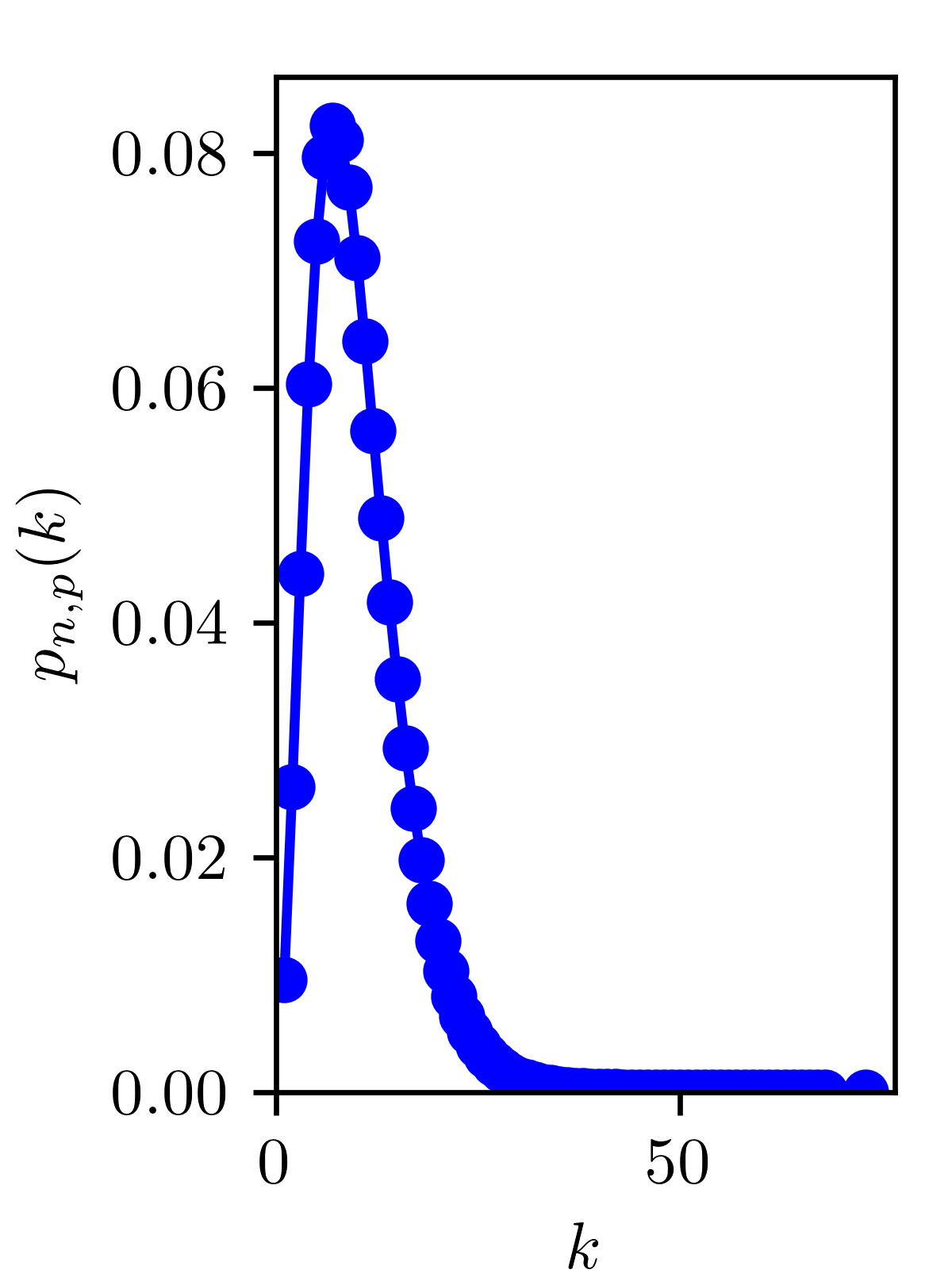}
\includegraphics[scale=0.95]{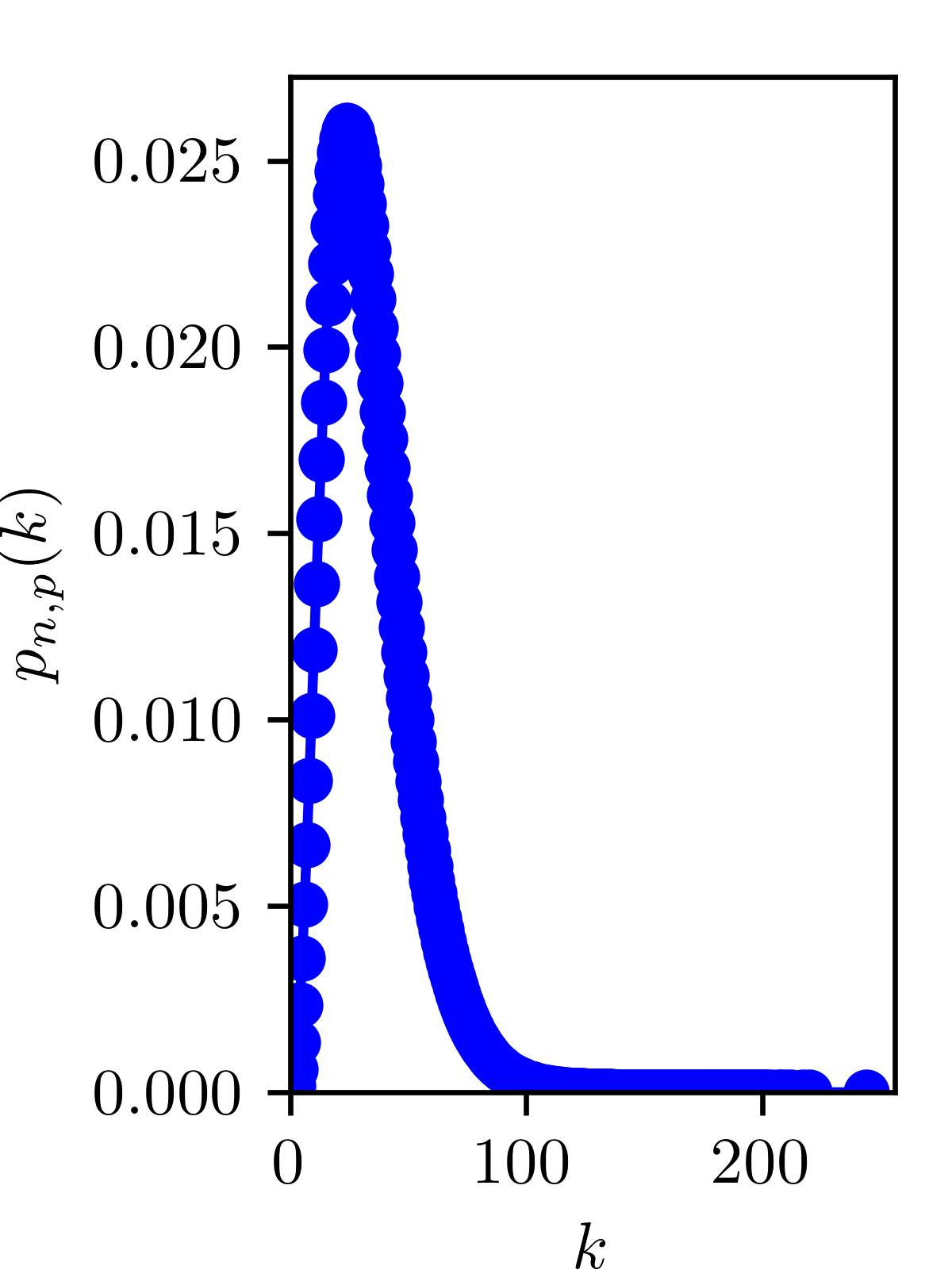}
\caption{The three degree distributions described in Table~\ref{table:degrees}.} 
\label{fig:degreedist}
\end{figure}

The parameters chosen are reported in Table~\ref{table:degrees}, and represent degree distributions of increasing variance, see figure~\ref{fig:degreedist}, moving from almost no variance (delta-like degree distribution) to a degree distribution with a longer tail, akin to a scale-free network. 

For all network-based models, we consider the simple susceptible-infected-recovered model (SIR) on a network. The infection is driven by a per-link transmission rate $\tau$ and a recovery rate $\gamma$. Mostly, we focus on the corresponding mean-field models with increasing level of accuracy in terms of what network features are incorporated.

\begin{table}[h!]
\centering
\begin{tabular}{|l|l|l|l|l|l|}
\hline
Name & $n$ & $p$  & $\langle k \rangle$ & $\sigma^2$ &$\tau$ \\ \cline{1-1}
\hline
delta-like                 & $1$   & $0.99$ & $10$                  & $1$  & $0.016$       \\
normal-like                & $3.86$  & $0.3$  & $10$                  & $30$  & $0.016$   \\
scale-free-like            & $1.07$  & $0.107$  & $10$                  & $300$  & $0.016$      \\
\hline
\end{tabular}
\caption{The three degree distributions considered. The delta-like distribution is shifted by $9$, as its mean would be $1$ otherwise. The reason for this choice is the fact that in negative binomials the variance cannot be lower than the mean. Normal-like and scale-free like distributions instead are shifted by one, so that the minimum degree is $1$. The resulting degree distributions are shown in figure~\ref{fig:degreedist}.}\label{table:degrees}
\end{table}

\subsection{Degree-based mean field model}
In the degree-based mean field model (also called Heterogeneous Mean Field~\cite{pastor2015epidemic}) we denote by $[S]_k(t)$ the expected number of susceptibles with degree $k$ at time $t$, similarly for $[I]_k$ and $[R]_k$. We define $[S] = \sum_{k=1}^\infty [S]_k$, similarly $[I]$ and $[R]$. The closure is made at the level of individuals, meaning that the infection pressure across a link is simply averaged across the entire spectrum of infected nodes. The resulting ODEs are
\begin{eqnarray}
 \dot{[S_k]} &=& -\tau k [S_k] \pi_I, \nonumber \\
 \dot{[I_k]} &=& \tau k [S_k] \pi_I - \gamma[I_k],\nonumber \\
 \dot{[R_k]} &=& \gamma[I_k],\nonumber \\
\pi_I &=& \frac{\sum_{\ell=1}^M \ell [I_\ell]}{\sum_{\ell=1}^M \ell N_\ell}, \label{eq:degree-based_mean-field}
\end{eqnarray}
where $N_\ell=P_{n,p}(\ell)N$ is the number of nodes with degree $\ell$. This system effectively keeps track of degree and heterogeneity in it, but mixing between nodes of different degrees happens at random but proportionally to degree~\cite{pastor2015epidemic,Kiss2017}, with clustering (the tendency of nodes to form connected triples) playing no role.

The degree-based mean field model can be derived exactly, under the assumption that individuals with degree $k$ reselect their partners very rapidly, so at every moment, the status of a node is independent of the status of its current partners.  In reality, many edges are long-lasting, and so correlations build up: a randomly selected infected node is more likely to connect to another infected node than a randomly selected susceptible node.  More complex models are needed to correct this.
\subsection{Heterogeneous pairwise without and with clustering}
In the heterogeneous pairwise model, we also count pairs: for example $[A_kB_\ell]$ is the expected number of links connecting a node of degree $k$ in state $A$ to a node of degree $\ell$ in state $B$~\cite{house2011insights,Kiss2017}; likewise for triples of the form $[A_kB_\ell C_m]$. The closure is done at the level of pairs (i.e. triples are approximated by singles and pairs), and hence an approximation for the triples are needed. These are given by 
\begin{equation}
[A_kB_\ell C_m] = \frac{\ell-1}{\ell}\left((1-\varphi)\frac{[A_kB_\ell][B_\ell C_m]}{[B_j]} + \varphi\frac{[A_kB_\ell][B_\ell C_m][C_mA_k]}{[A_k][B_\ell][C_m]}\right),    
\label{eq:triple-closure}
\end{equation}
where $\varphi$ is the global clustering coefficient in the network. For the un-clustered case we simply set $\varphi=0$.
The resulting ODEs are,
\begin{eqnarray}
 \dot{[S_k]} &=& -\tau \sum_{\ell}[S_kI_{\ell}], \nonumber \\
 \dot{[I_k]} &=& \tau \sum_{\ell}[S_kI_{\ell}] - \gamma[I_k],\nonumber \\
 \dot{[R_k]} &=& \gamma[I_k],\nonumber \\
    \dot{[S_kI_\ell]} &=& - \gamma[S_kI_\ell] + \tau \left(\sum_{\alpha}[S_kS_\ell I_{\alpha}] -\sum_{\alpha}[I_{\alpha}S_kI_\ell] -[S_kI_\ell] \right), \nonumber \\
 \dot{[S_kS_\ell]} &=& - \tau\left( [S_kS_\ell I] + [I S_k S_\ell] \right),  \nonumber\\
  \dot{[I_kI_\ell]} &=& -2\gamma[I_kI_\ell]+ \tau\left( \sum_{\alpha}[I_{\alpha}S_kI_\ell] + \sum_{\alpha}[I_kS_\ell I_{\alpha}]+[S_kI_\ell]+[I_kS_\ell] \right), 
 \label{eq:pairwise}
\end{eqnarray}
where triples are closed using equation~\eqref{eq:triple-closure}. The system  can be significantly simplified for the $\varphi=0$ case. When $\varphi>0$, the closures become more complicated and present further challenges when implemented numerically (see notes in Appendix~\ref{sec:het-PW-eq}).

The number of equations in the heterogeneous pairwise model grows very large if the network has degrees of many different types (e.g., because there is an equation for $\dot{[S_kI_\ell]}$ for every $k$, $\ell$ pair).  Generalizing this to more complex structures can become unwieldy. Edge-based compartmental models provide an alternative and are discussed next.

\subsection{Edge-based compartmental model with household structure and community transmission}
\begin{figure}[h!]
    \centering
    \includegraphics[trim={6cm 2cm 7cm 1cm},clip,scale=0.5]{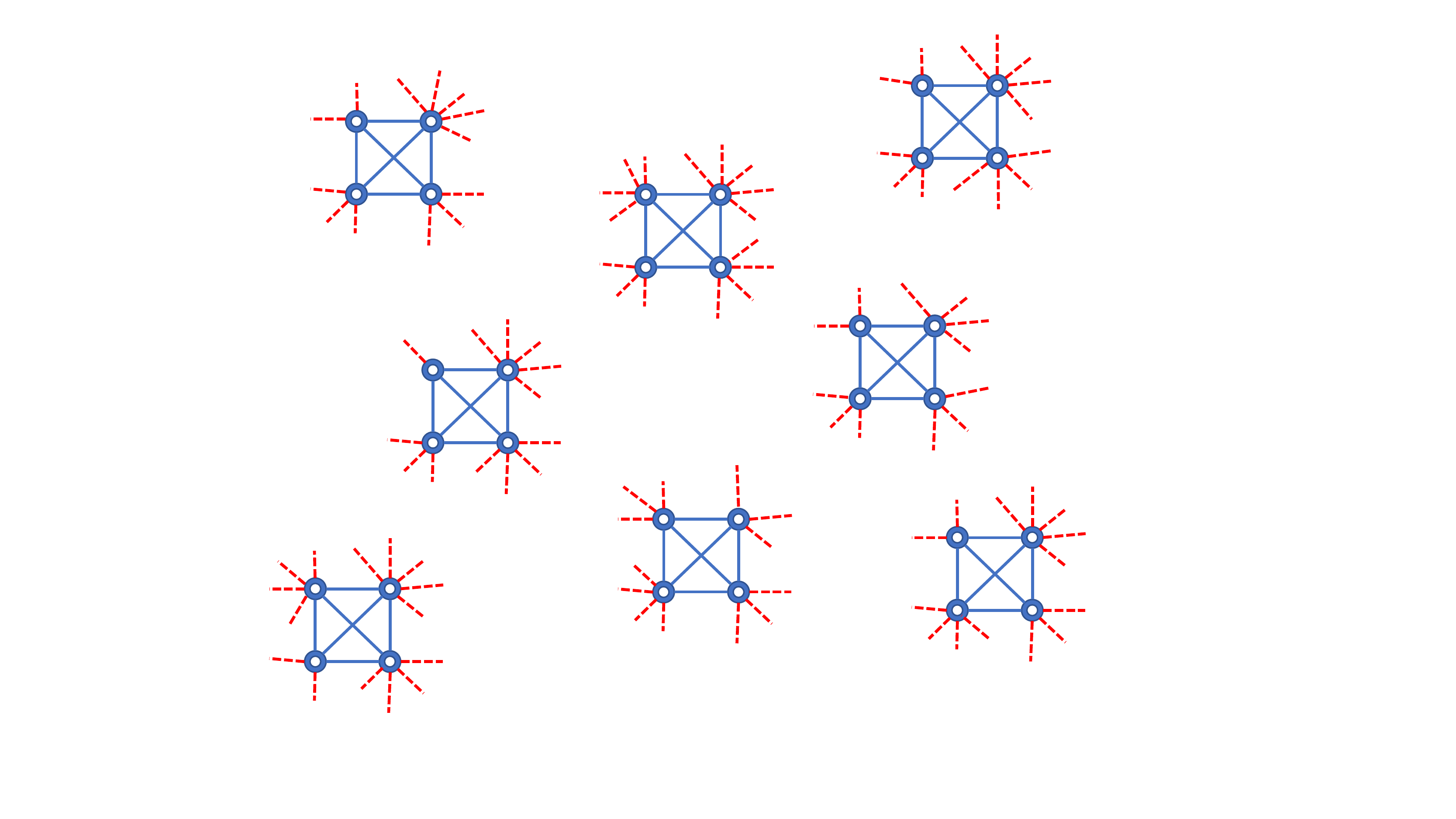}
    \caption{Caricature of the network model with households of size four and community stubs due to be connected up following the configuration model.}
    \label{fig:joel-net}
\end{figure}

It is interesting to consider models which explicitly distinguish between links that happen within the households and those that happen elsewhere, as lockdowns act mostly on inter-household contacts.
To consider household structures, we take advantage of the edge-based compartmental modelling (EBCM) framework~\cite{miller:volz,miller:ebcm_overview}, adapting the model in~\cite{volz:clustered_result} to build a model that (i) has a more realistic contact structure with households, and (ii) can distinguish between within household and community transmission, see figure~\ref{fig:joel-net}.  This model keeps the number of equations tractable.

We assume that individuals are divided into households of size $4$. Within households, there is complete mixing.  In addition, each individual has a number of contacts outside the household, which allow for community transmission. The equations for this model are given in Appendix~\ref{sec:ebcm-eq}.

\subsection{Age-structured compartmental model}
We use a version of the SEIRD compartmental model by Alvarez et al.~\cite{alvarez2015new} adapted to remove any built-in control measures (originally modelled as a Hill repression function modulating the number of daily contacts in response to control measures) and to include age-structured interactions in the population. The model is as follows: 

\begin{eqnarray}
\dot{S_{i}} &=& -\beta S_{i} \sum_{j=1}^{n} \mathbf{C}_{i j} I_{j} / N_{j}, \nonumber \\
\dot{E_{i}} &=& +\beta S_{i} \sum_{j=1}^{n} \mathbf{C}_{i j} I_{j} / N_{j}-\gamma_{E} E_{i}, \nonumber \\ 
\dot{I_{i}} &=& \gamma_{E} E_{i}-\gamma_{I} I_{i}, \nonumber \\
\dot{R_{i}} &=& +\left(1-m_{i}\right) \gamma_{I} I_{i}, \nonumber \\
\dot{D_{i}} &=& +m_{i} \gamma_{I} I_{i}, \label{alvarezeq}
\end{eqnarray}

where $\beta$, $\gamma_E$ and $\gamma_I$ are age-independent parameters denoting infectivity, rate at which exposed individuals become infected (inverse of incubation period) and rate at which infected individuals  recover or die (inverse of disease duration) respectively (note that $\gamma_I$ in this model corresponds to $\gamma$ in the above models). The $m_i$ are age-dependent mortality probabilities and control the fraction of those infected individuals who die. Susceptible individuals become exposed proportionally to a force of infection defined as the product of the contagion matrix with the prevalence by age. The contagion matrix is simply the product of the intrinsic infectivity of the epidemic and the daily contacts of individuals in age group $i$ with individuals from age group $j$, $\mathbf{C}_{ij}$. Finally, $n$ is the number of age groups considered and $N_i$ is the count of individuals in age group $i$. 

\subsection{Epidemic parameters}\label{sec:R0}
Ling et al.~\cite{Ling2020} reported that the median time from symptoms onset to first negative RT-PCR in oropharyngeal swabs of convalescent patients was around $10$ days, and further evidence~\cite{Wei2020} suggested that pre-symptomatic infection could happen $1-3$ days before the first symptoms appear. Accordingly, we set $\gamma = 1/14$ (i.e. an average of two weeks before recovery). Before setting $\tau$, we summarise the expression for $R_0$ for the various models that we consider. For the heterogeneous degree-based mean-field model~\cite{Kiss2017} we have
\begin{equation}
R_{0}=\frac{\tau \langle k^2 \rangle}{\gamma \langle k \rangle}.
\end{equation}
For the heterogeneous pairwise model, we use
\begin{equation}
R_0 =  \frac{\tau}{\tau+\gamma}\frac{\langle k^2 \rangle -\langle k \rangle}{\langle k \rangle}. 
\end{equation}
For the edge-based model, we set the in-household infection parameter $\beta_h$ to be $3-5$ times bigger than the inter-households infection parameter $\beta_c$. 

The basic reproduction number of the edge-based and the age structured models are given by the leading eigenvalues of the following two next-generation matrices~\cite{diekmannConstructionNextgenerationMatrices2010}. First, for the edge-based compartmental model, based on~\cite{pellis2012reproduction}, we have
\begin{equation}
A=\begin{bmatrix}
\tilde{\mu}_c \mu_0 & 1 & 0 & 0\\
\mu_c \mu_1 & 0 & 1 & 0\\
\mu_c \mu_2 & 0 & 0 & 1\\
\mu_c \mu_3 & 0 & 0 & 0
\end{bmatrix},
\end{equation}
where 
\begin{equation}
\tilde{\mu}_c=\frac{\beta_{c}}{(\beta_{c}+\gamma)}E[\tilde{D}],\,\,\, \mu_c=\frac{\beta_{c}}{(\beta_{c}+\gamma)}E[D],
\end{equation}
with $\tilde{D}$ and $D$ the distribution of the excess degree (i.e. the distribution of the left-over edges attached to a node reached by following one random edge, see~\cite{newman2018networks}) and degree distribution of the network, respectively. $\mu_0$, $\mu_1$, $\mu_2$ and $\mu_3$ are the expected number of infected in generation 0, 1, 2 and 3 in a household of size 4. These values can be found in~\cite{britton2019stochastic} on page 222/223, with $\varphi_{I}(\beta_h)=\gamma/(\beta_h+\gamma)$.

Finally, the $R_0$ for the age-structured compartmental model is given by the largest eigenvalue of: 
$$\frac{\beta}{\gamma^I} \left[\begin{array}{cccc}
\frac{N_1}{N_1}\mathbf{C}_{11} & \frac{N_1}{N_2}\mathbf{C}_{12} & \cdots & \frac{N_1}{N_n}\mathbf{C}_{1n} \\
\frac{N_2}{N_1}\mathbf{C}_{21} & \frac{N_2}{N_2}\mathbf{C}_{22} & \cdots & \frac{N_2}{N_n}\mathbf{C}_{2n} \\
\vdots & \vdots & \ddots & \vdots \\
\frac{N_n}{N_1}\mathbf{C}_{n1} & \frac{N_n}{N_2}\mathbf{C}_{n2} & \cdots & \frac{N_n}{N_n}\mathbf{C}_{nn} \\
\end{array}\right].$$

where $\beta$ is the intrinsic infectivity, $\gamma^I$ is the rate at which infected individuals either recover or die, $N_i$ is the size of the population in age group $i$, and $n = 18$ is the number of age groups in the model. $\mathbf{C}$ is the age-mixing matrix and the normalisation factor $\frac{N_i}{N_j}$ comes from the fact that at $t_0$, there are only susceptible individuals in the population of each age group and therefore $S_i = N_i$ in the partial derivative with respect to $I$ of the r.h.s of the second equation in system~\eqref{alvarezeq}. 


\section{Results}\label{sec:results}
Most of our scenarios are concerned with determining the impact of model and demographic heterogeneities on the DIHI levels. In well-mixed homogeneous populations, each individual contributes equally to spreading, and therefore DIHI is a well-defined quantity, independent of whom has been infected during the first wave. In models with degree heterogeneities, nodes with higher degree contribute to the spreading of the disease much more than nodes with fewer links. This means that depending on whom has been infected, we can observe different levels of DIHI. This effect, however, does not show in our results since the DIHI is based on the infections accumulated during the first wave as determined by our equations.

It is well known that in the simple compartmental model herd immunity at time $t=0$ is achieved as long as at least $1-1/R_0$ of the susceptible individuals are removed or immunised. In line with~\cite{Britton2020}, our general setup is that an initial epidemic spreading freely for a short time is intervened upon by implementing a lockdown period of fixed duration. Afterwards, all parameters immediately return to their pre-lockdown values. In the most basic case this is done by keeping the network or the mixing matrix the same and multiplying the transmission rate by a constant $0<\alpha<1$. Crucially, however, we also explore the implications of how the lockdown is modelled, that is, we investigate the difference between reducing the transmission rate whilst keeping the network the same and changing the contact network, the latter being more in tune with what happens in reality. In the more realistic age-structured model, we compare a reduction of all entries in the mixing matrix with a number of scenarios involving school closure and work distancing.

For the edge-based compartmental model we focus on final epidemic size but still under the assumption of a lockdown period. This is because we want to compare how the two different strategies affect the eventual outcome of the epidemic, rather than how the optimal $\alpha$ varies between the two strategies, a comparison that would be difficult to interpret. 

\begin{figure}[]
    \centering
    \includegraphics[scale=0.8]{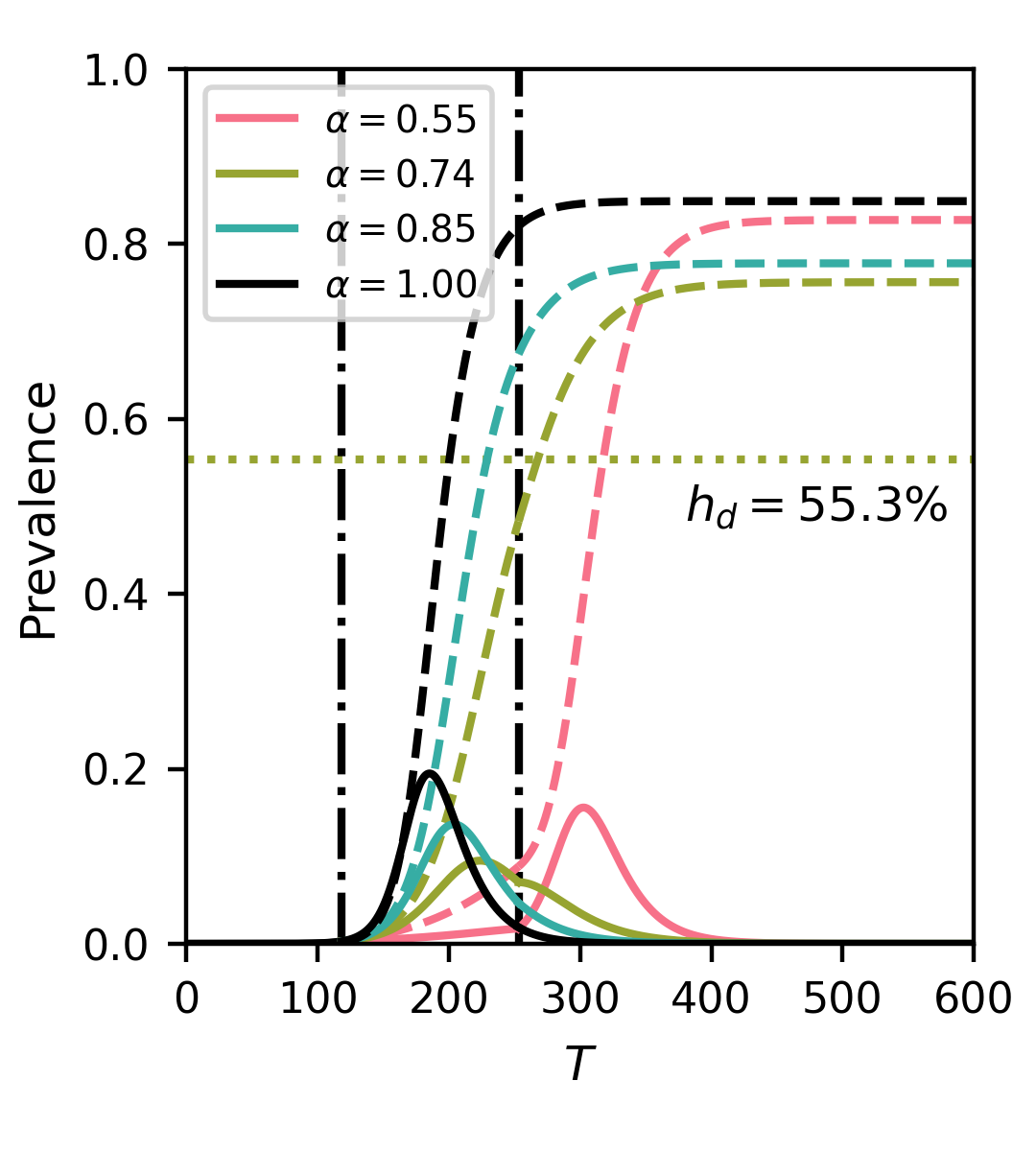}
    \includegraphics[scale=0.8]{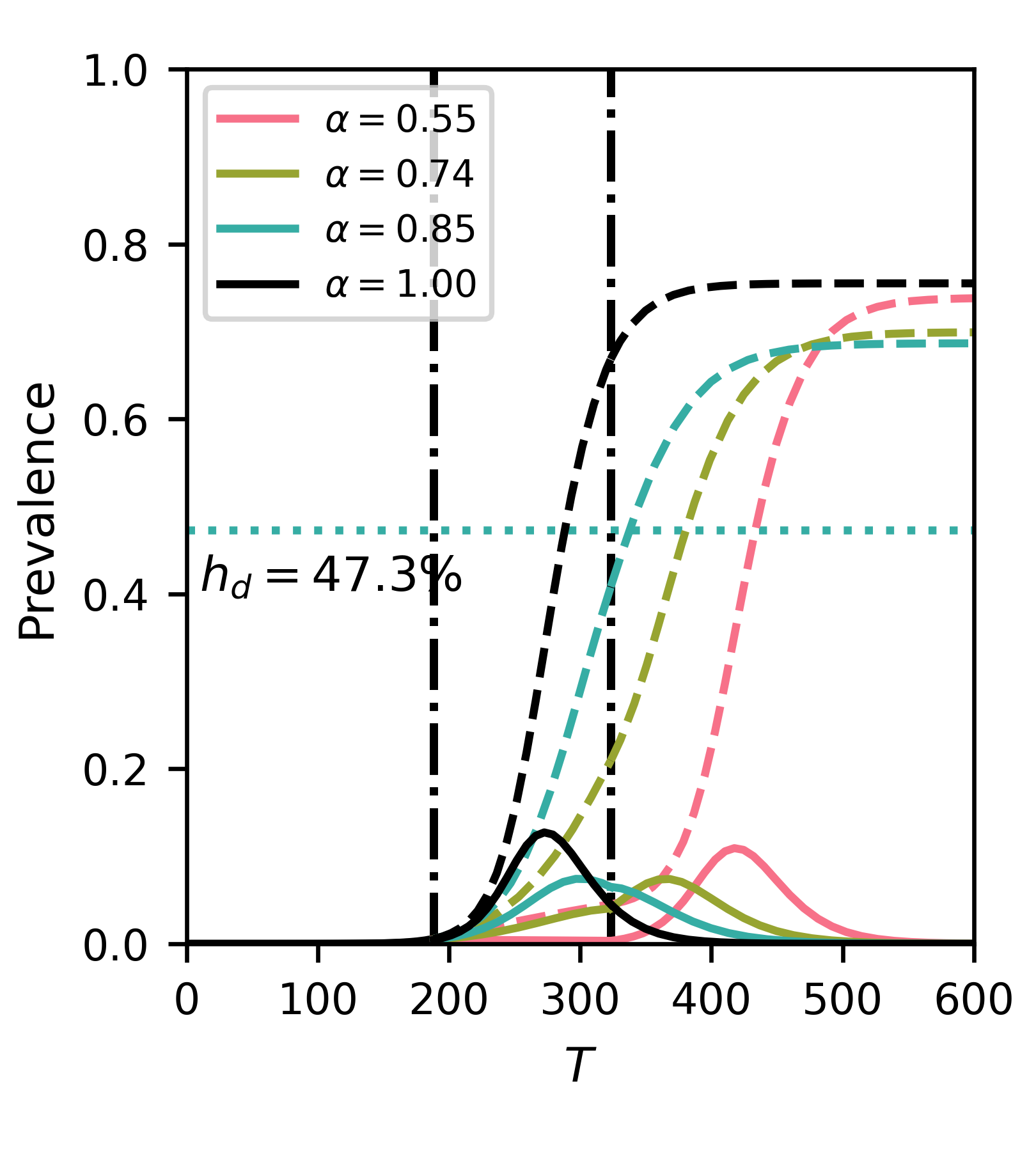}    
    \includegraphics[scale=0.8]{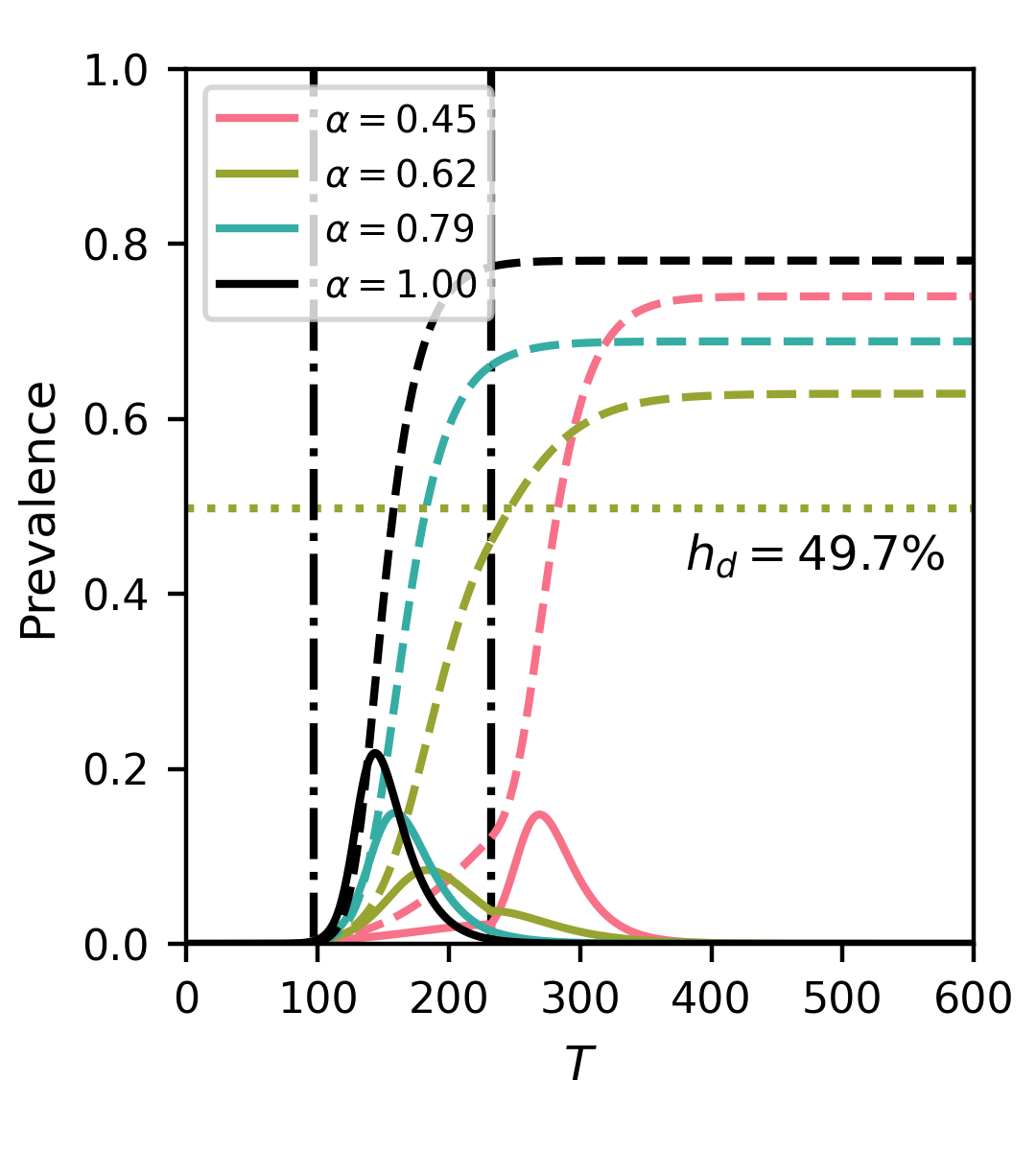}
    \includegraphics[scale=0.8]{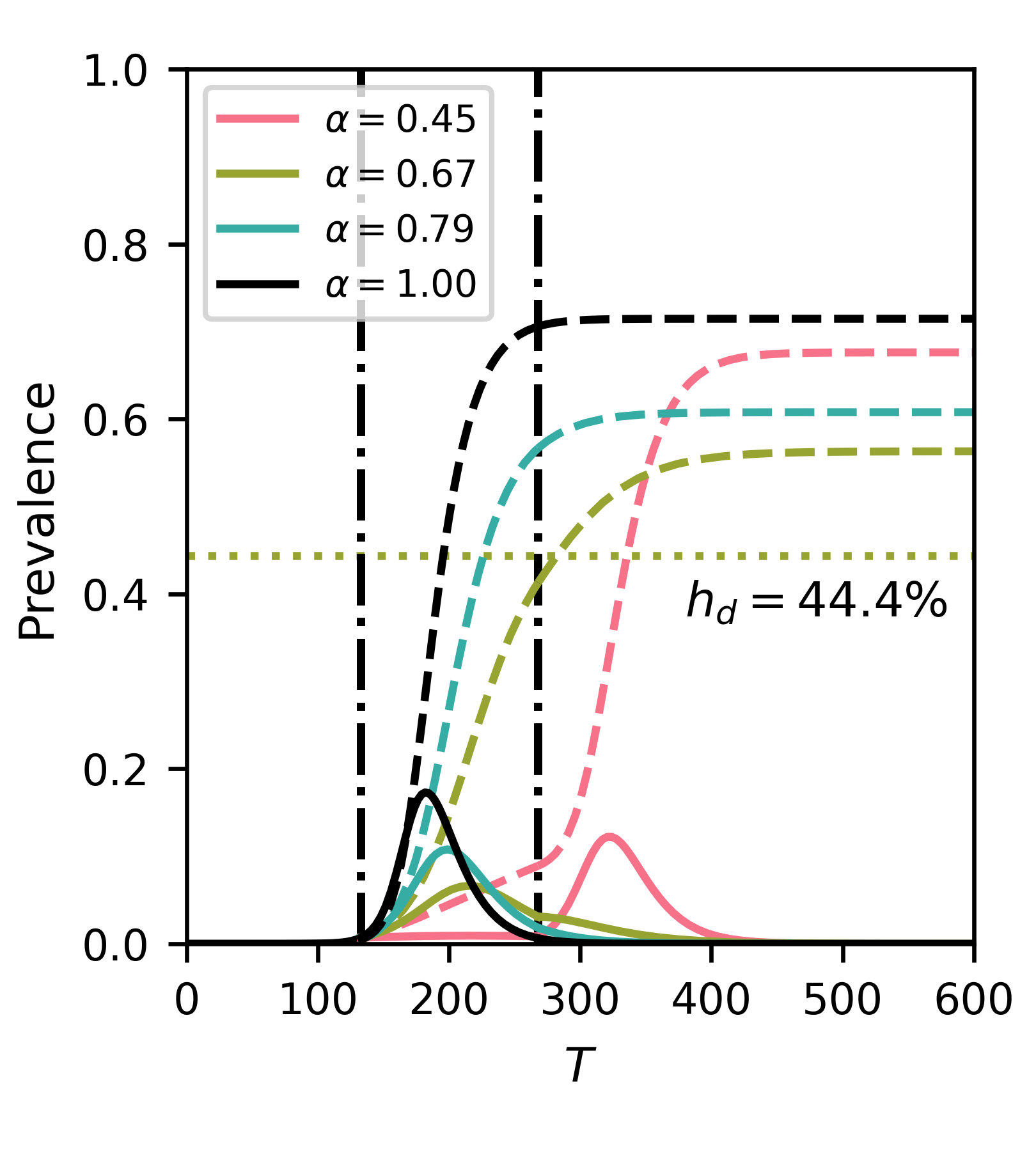}
    \includegraphics[scale=0.8]{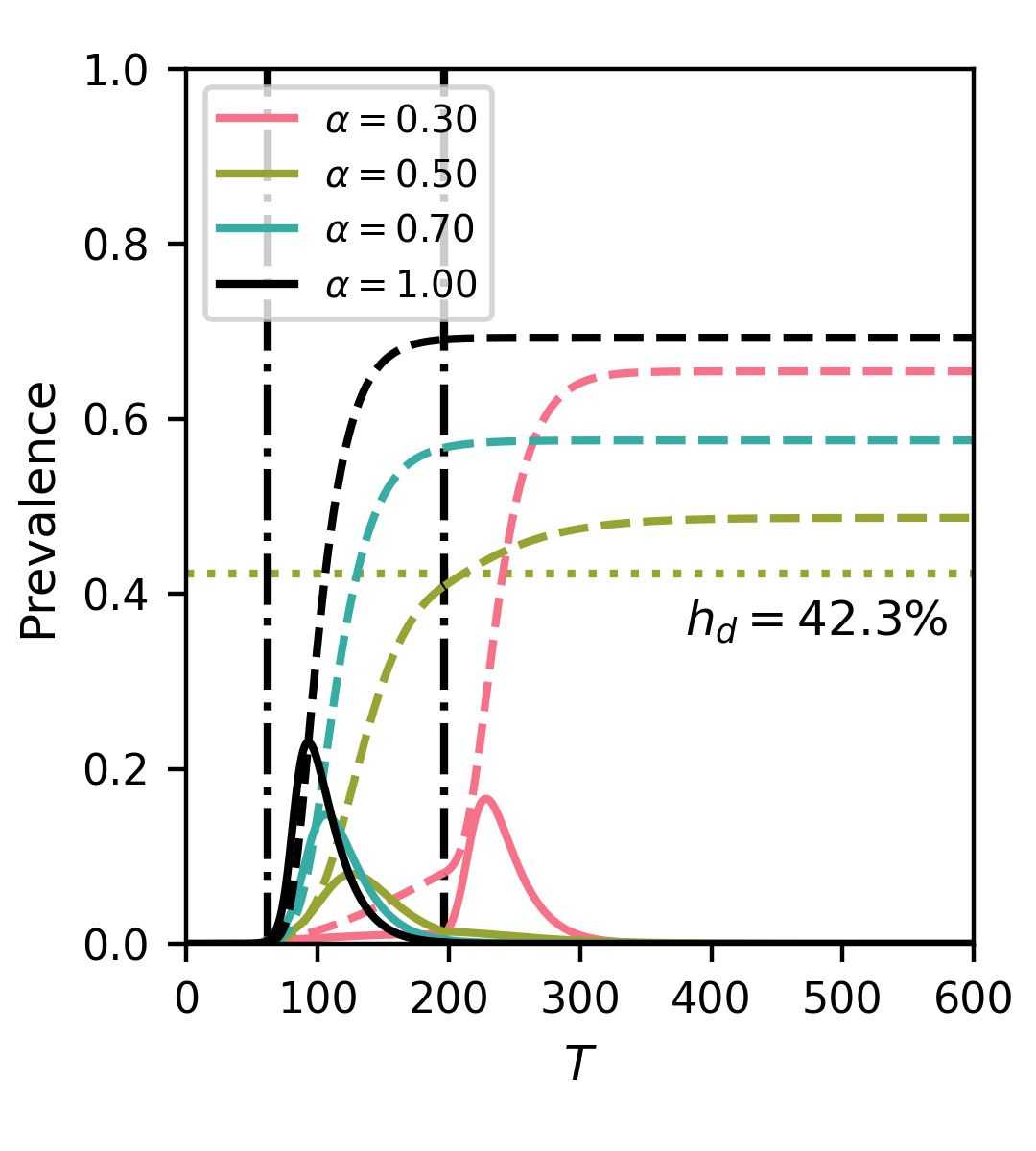}
    \includegraphics[scale=0.8]{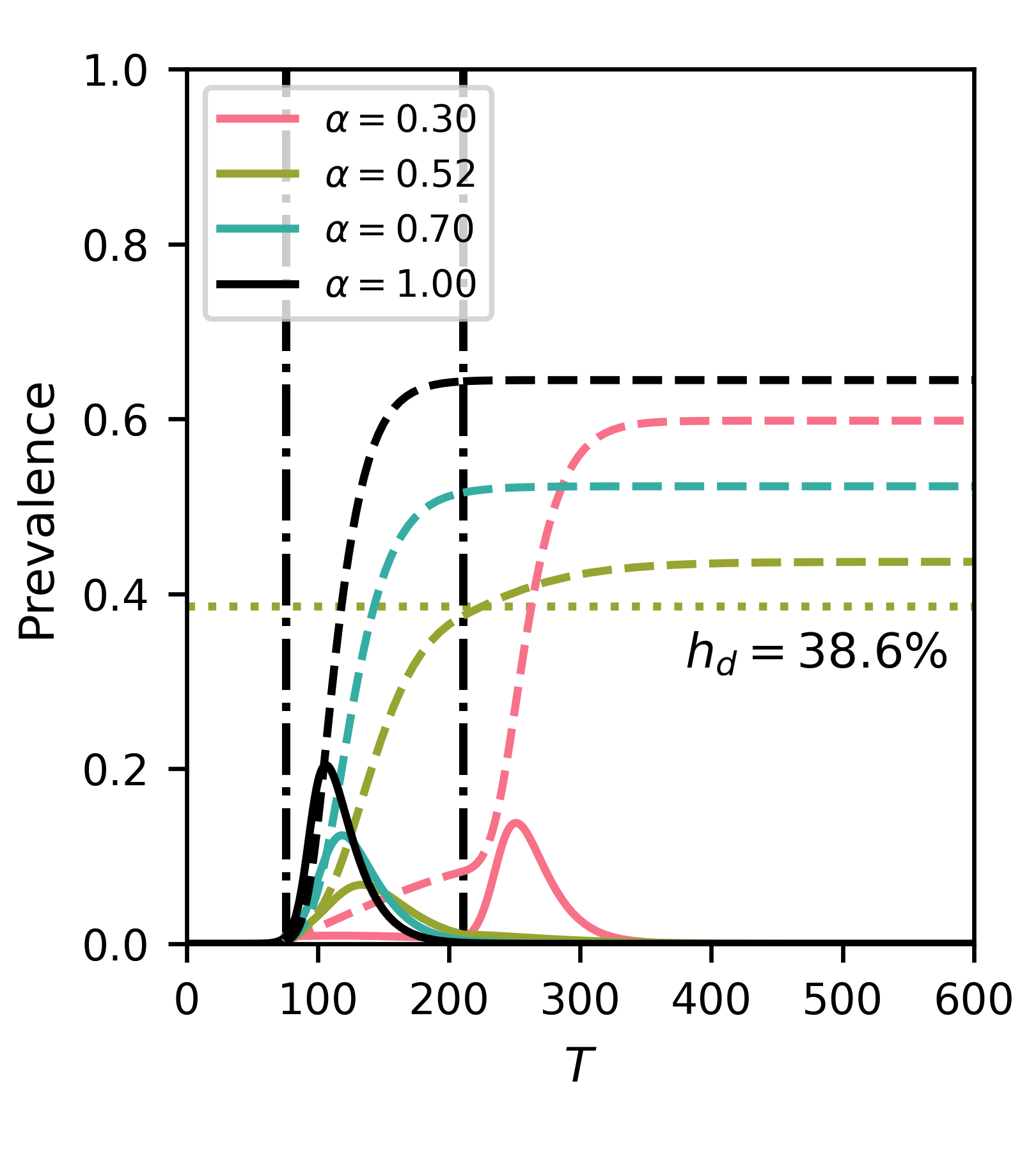}    
    \caption{Optimal $\alpha$ (see legend) and DIHI in delta-like (first row), normal-like (second row) and scale-free-like (third row) networks using the heterogeneous mean-field (first column) and heterogeneous pairwise model with $\varphi=0$ (second column). Continuous curves indicate $[I](t)$, while dashed curves indicate $[R](t)$. The two vertical curves represent the beginning and the end of the lockdown. Duration of lockdown is $130$ days. Finally, the horizontal line and the corresponding percentage reported are the cumulative prevalence at the end of lockdown for the best strategy that does not allow for a second wave.}
    \label{fig:epidemics}
\end{figure}

There is an extremely strong relation between  the speed or rate of spread of the uncontrolled epidemic and the timing and length of the lockdown. In fact this can be visualised in terms of the `flattening of the curve' argument. A reduction of the transmission rate during lockdown leads to a flattening of the epidemic curve with a reduced peak and an extended duration which ideally should fit within the control period. This means that if the epidemic grows quickly and the lockdown period is short two outcomes are possible. First, a fast growing epidemic with a short lockdown period needs to be met with a significant reduction in the transmission rate, i.e., small values of $\alpha$. This will lead to a reduced epidemic which does not have enough time to unfold and the lifting of lockdown is followed by a full blown epidemic. Second, if the reduction is not strong enough (i.e., larger values of $\alpha$), then a significant epidemic will occur during the lockdown itself with no further peak after lifting control~\cite{di2020timing}, see also figure~\ref{fig:epidemics}.


\subsection{Contact heterogeneity and clustering}\label{sec:results_DB_PW}
The main focus here is to investigate the impact of degree-heterogeneity and clustering on herd immunity induced by a first wave of the epidemic. In networks with heterogeneous degrees, one can preferentially target high-risk individuals and this will reduce the number needed to achieve DIHI. Such heterogeneity can be exploited in many ways: for example, targeted immunisations~\cite{albert2000error} and acquaintance  immunisation~\cite{cohen2003efficient}. Equally, the epidemic itself typically finds the high-risk groups first and thus `removes' important individuals or risk groups. In line with~\cite{Britton2020,gomes2020individual}, we exploit this fact and consider different levels of degree-heterogeneity and different mean-field models to explore what happens in the wake of the lockdown period when some level of spreading is possible.

In each scenario considered, we seeded nodes with degree $k=10$ with $[I_k](0)= 5$ infected individuals, the rest of the population being fully susceptible. We let the epidemic run until the cumulative number of infected people reached $0.5\%$ of the population. Then, a one-shot intervention lasting exactly $T=130$ days kicked in. During lockdown,  control measures made $\tau \to \tilde{\tau}_0 = \alpha \tau$ (by acting on $\tau$). Afterwards, lockdown was lifted and $\tau$ immediately returned to its pre-lockdown value. For the edge-based-compartmental and age-structured models, lockdown also involved preferential interventions on community or household links and modulation of the mixing matrix, respectively.

Figure~\ref{fig:epidemics} shows results from the degree-based mean field model (left column) and heterogeneous pairwise model without clustering (right column) for networks with increasing levels of degree heterogeneity (from top to bottom). In each case, we find the optimal $\alpha$ (a simple down scaling of the transmission rate without change to the network) and report the number of infections required to achieve DIHI  (i.e., total of infected and recovered nodes at the end of lockdown such that the the epidemic after lockdown is subcritical). Several observations can be made. First, for both models, aggressive control (value of $\alpha$) leads to a second wave. Equally, if the control is too weak (high value of $\alpha$) the epidemic will still run its course during the first wave with some reduction in the final size. Hence, there is an optimal value of $\alpha$ for which the final epidemic size is smallest and the epidemic post-lockdown is subcritical.

Both models clearly show that DIHI decreases with variance. Despite displaying the same overall trends, the two models are quantitatively different. In a like-for-like comparison, the degree-based heterogeneous mean-field model leads to larger overall epidemics. This is to be expected since this model does not keep track of the links explicitly and thus over estimates what would happen in a true simulation on an explicit network~\cite{Kiss2017}. The pairwise model, however, accounts for links and correlations and leads to epidemics that are typically less potent. Furthermore, the heterogeneous pairwise model leads to smaller values of DIHI showing that the accuracy with which the network structure is accounted for matters. This demonstrates that model choice is important as the precise levels of DIHI matter in a real-world scenario.

\begin{figure}[h]
      \includegraphics[scale=0.86]{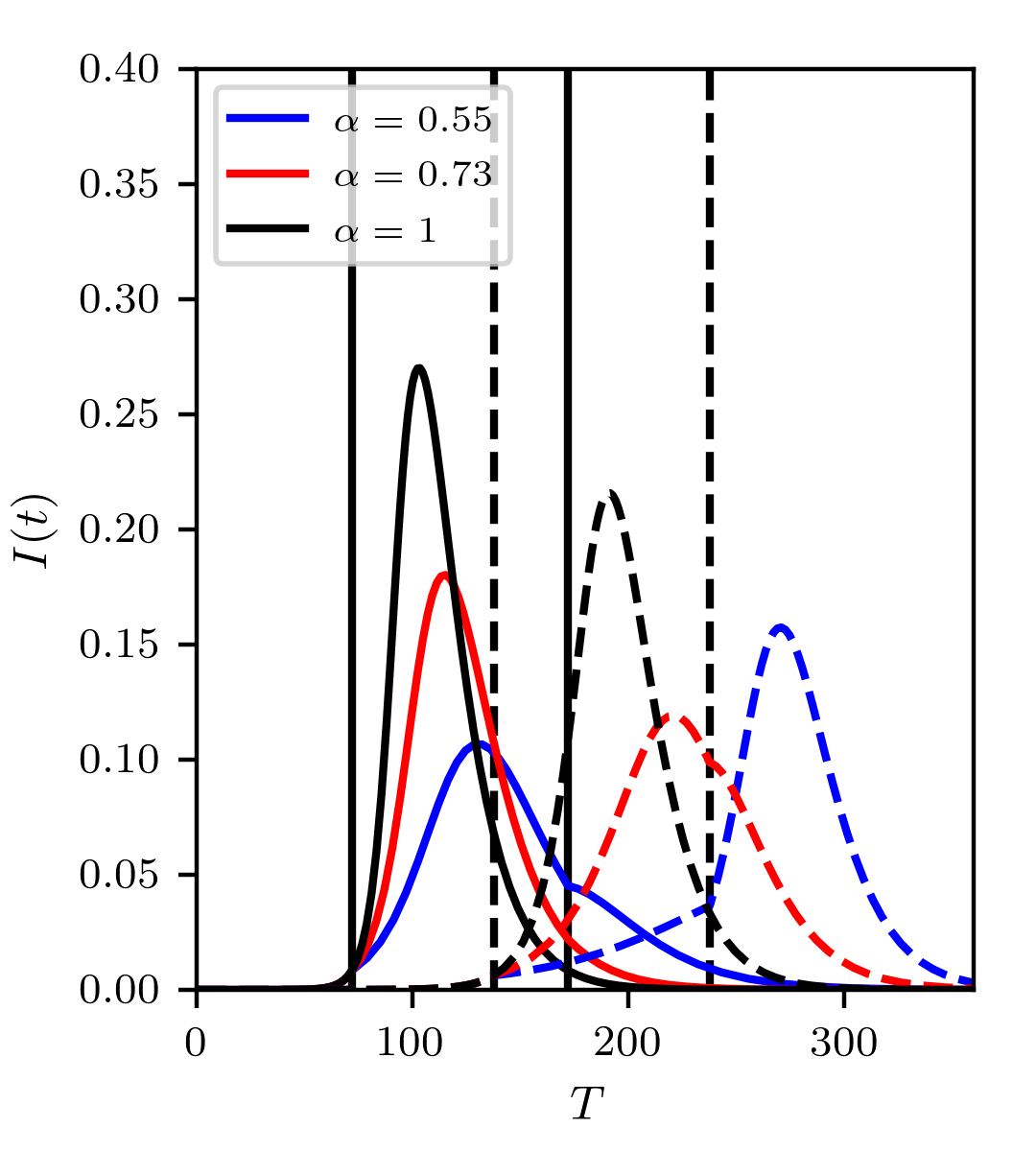}\centering
    \includegraphics[scale=0.86]{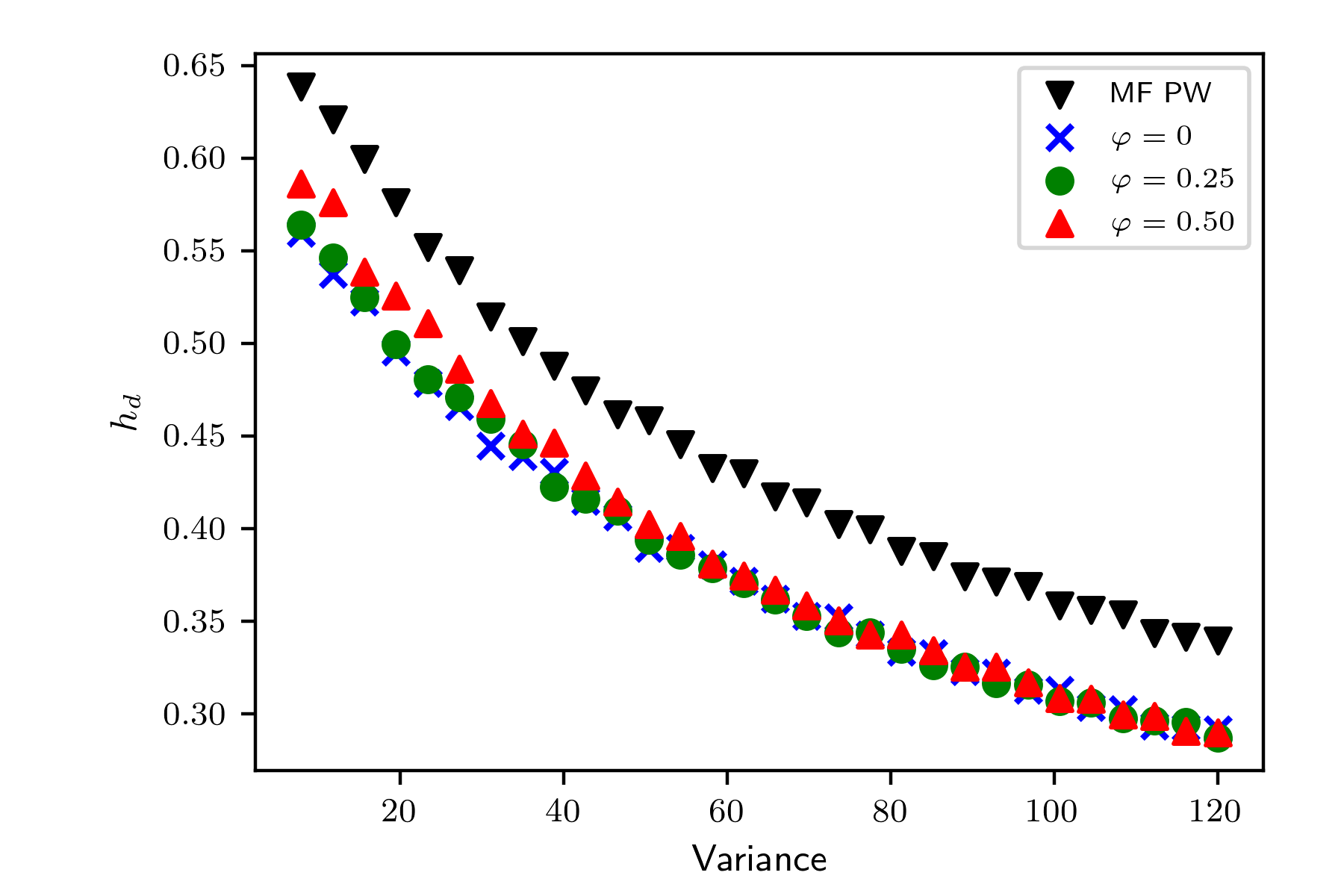}
\caption{(left)  Difference between control acting on un-clustered networks (continuous lines) and clustered networks (dashed lines), with clustering coefficient $\varphi=0.5$, corresponding to the second point on the x-axis of the right panel. Vertical lines are at the beginning (continuous) and end (dashed) of control. Blue curve is optimal control for $\varphi=0$, red for $\varphi=0.5$. (right) Impact of variance in degree distribution on DIHI $h_e$, for different pairwise models with different values of $\varphi$. Average degree is $\langle k \rangle = 6$, $\tau = 0.04$ and $\gamma =1/14$. Control duration is 100 days from the moment $I(t)+R(t)\geq 0.025$. 
}
\label{fig:hetpwvariance}
\end{figure}

The effect of clustering is illustrated by figure~\ref{fig:hetpwvariance}.
Typically, clustering lengthens the duration of the epidemic and lowers the peak when compared to the unclustered case (see also ~\cite{volz:clustered_result}). The final epidemic size is also smaller. This means that there is a small amount of leeway for implementing control and that the control effort can be smaller compared to the unclustered case.  It is worth noting that the final epidemic size is also smallest at the optimal $\alpha$ value (see also~\cite{Britton2020}).

Finally, opting for the more accurate heterogeneous pairwise model, the level of DIHI is plotted for increasing values of variance and for different clustering levels, see figure~\ref{fig:hetpwvariance}. It is clear that higher variance can drive DIHI levels to as low as $30\%$. Clustering in the heterogeneous pairwise model leads to even smaller values of the DIHI, although increasing variance in the degree negates the effect of clustering and the DIHI levels are very similar to those observed with $\varphi=0.25$ and $\varphi=0.5$. This highlights the non-trivial interactions between network properties where clustering has biggest impact in sparse networks and where high levels of degree heterogeneity can negate the effect of clustering.

\subsection{All versus community control only}\label{sec:results_EBCM}
Although the previous models do account for some important contact network features, they are not ideal to capture structure such as households. Being able to capture households explicitly and having the flexibility to differentiate between household and community transmission is important because many of the interventions available to us (e.g., closing schools and workplaces) affect community transmission differently from household transmission. Thus, a distinctive feature of most lockdown measures is a change in network structure, rather than a global reduction in transmission rate.

Although our model is not an exact reflection of true population structure, it allows us to investigate whether an intervention that disproportionately affects between-household transmission can be appropriately captured by a model that treats the intervention as reducing all transmission rates.


\begin{figure}
    \centering
    \includegraphics{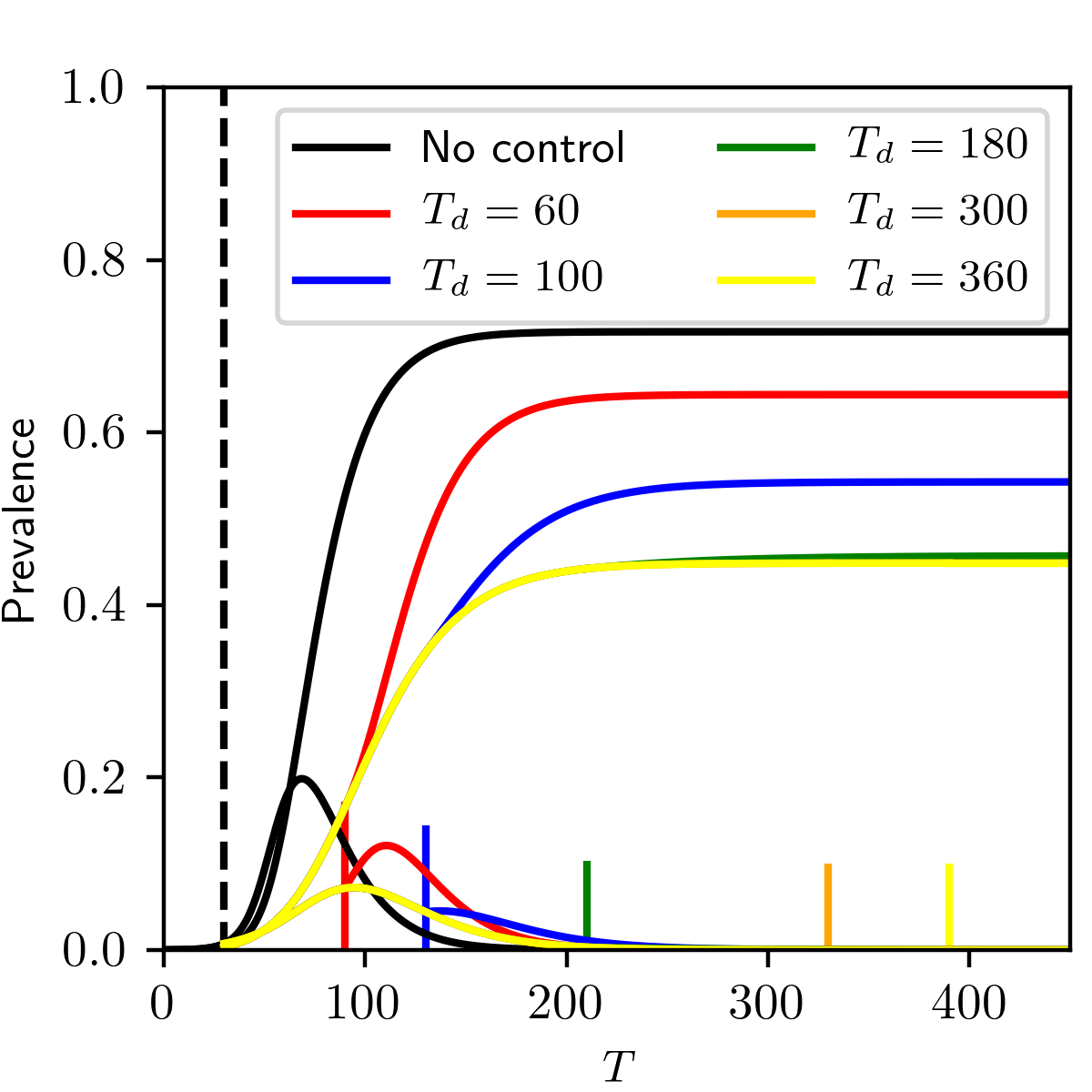}
    \includegraphics{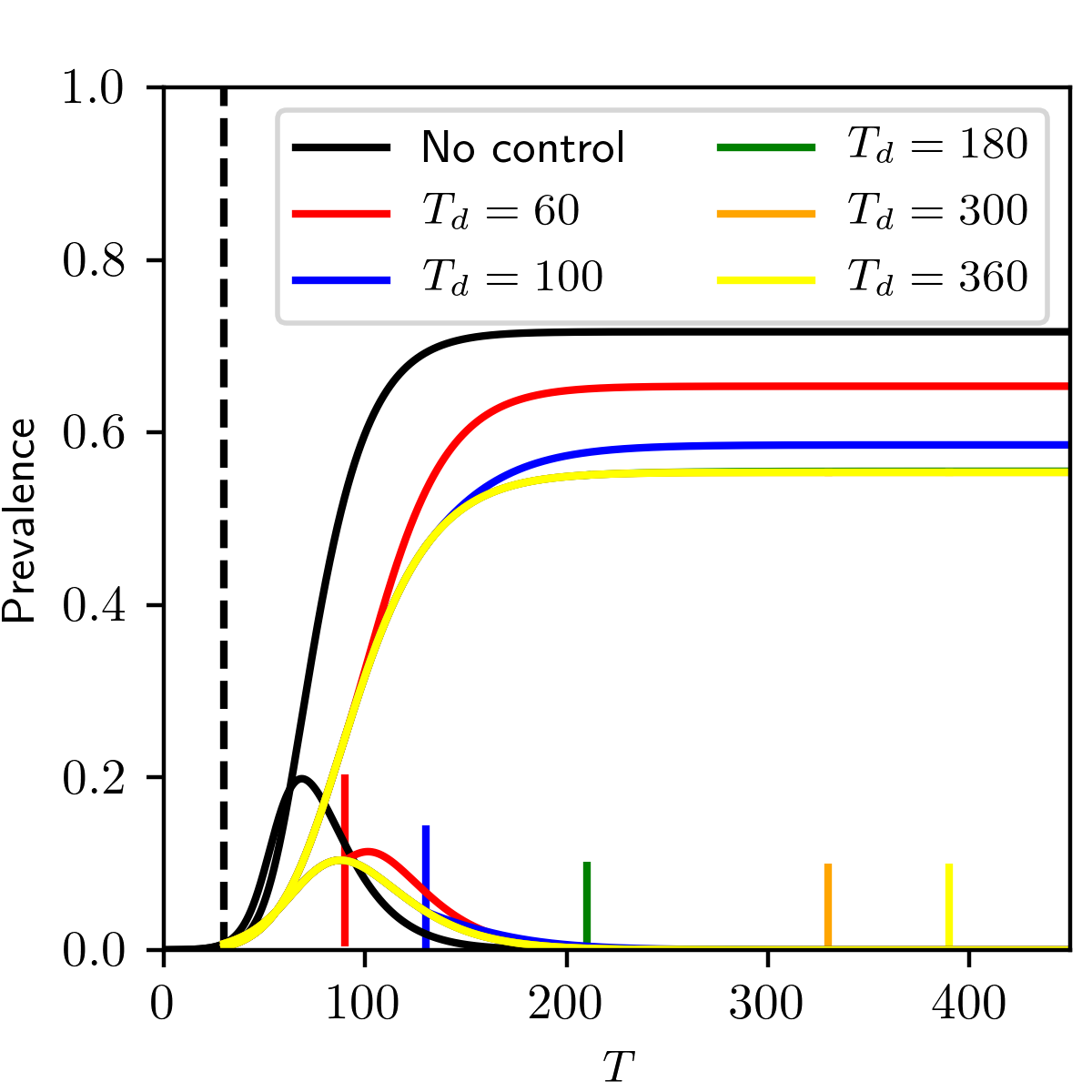}
    \caption{Control scenarios based on the EBCM model with intervention scaling factor of $\alpha=0.6$ starting at $T = 60$ (dashed vertical line), and lasting for different durations (continuous vertical lines). (left) Intervention on the whole network, (right) intervention on the community structure. Parameters of the epidemic and community network are $\langle k \rangle = 4, \sigma^2 = 7.5, \beta_h=0.045, \beta_c=0.015$.}
    \label{fig:duration_impact}
\end{figure}
\begin{figure}
    \centering
    \includegraphics{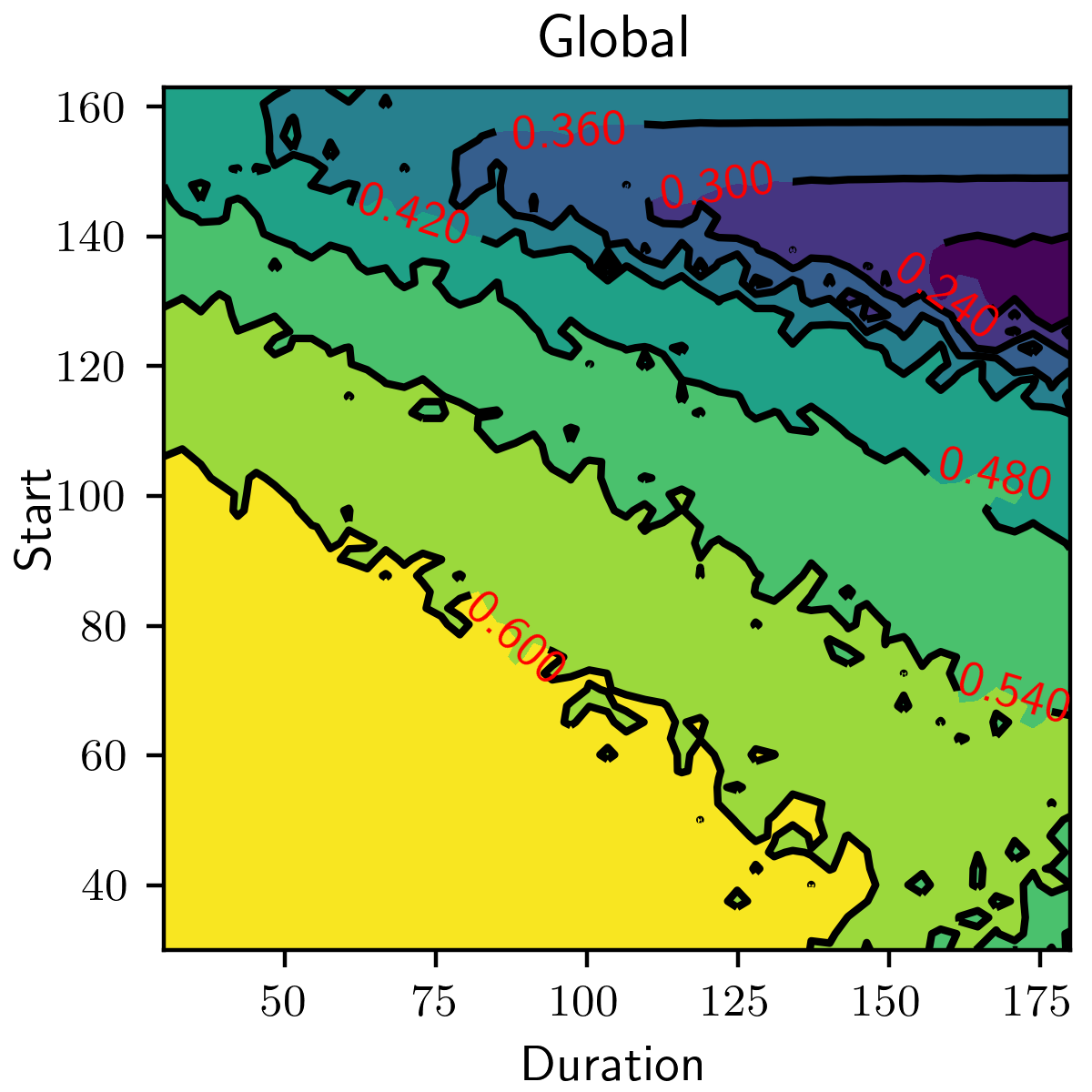}
    \includegraphics{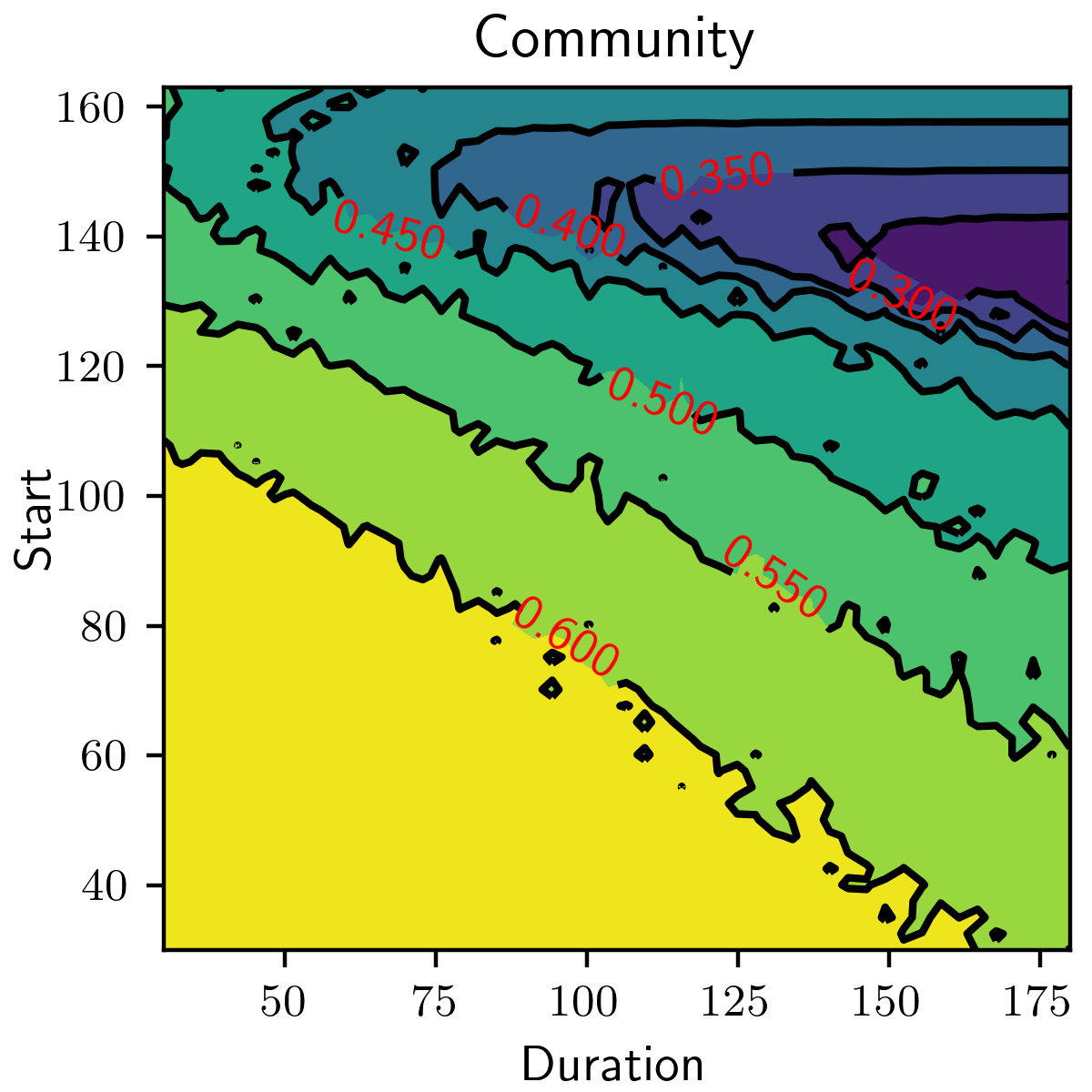}
    \caption{Final epidemic sizes based on the EBCM  as a function of the beginning of lockdown and its duration, with two different strategies: intervention on the whole network (left) or intervention on the community links only (right). Each value is the minimum final epidemic size that can be obtained for varying $\alpha$. Parameters of the epidemic and community network are: $\langle k \rangle =4$, $\sigma^2 = 7.5$, $\beta_h =0.045, \beta_c = 0.015$}
    \label{fig:contourplot}
\end{figure}

 For a given start time and intervention strength, if the duration of the intervention is long enough, the number of infections becomes very small and the eventual rebound has the same shape, regardless of duration.  The optimal intervention strength leaves the population sitting at the DIHI threshold when the epidemic dies out.  When the intervention is lifted, no rebound occurs.  Thus the outcomes of the optimal intervention are the same if the duration is long enough.  In figure~\ref{fig:duration_impact} and at the top right of figure~\ref{fig:contourplot} we notice that if the intervention is long enough, the optimal final size  becomes independent of duration.  To see this for earlier start times requires longer durations. 
 Our argument to explain why this happens is as follows: suppose that for a given $\alpha$, we find the smallest time such that DIHI is achieved. Prolonging the intervention for longer than said time will result in the number of actively infectious people reducing even further by end of lockdown but not change the downward trend of the epidemic, which will eventually die out. 

Comparing the top right of both panels in Figure~\ref{fig:contourplot} show that the optimal community intervention allows more infections than the optimal global intervention.  

The intuition behind this is based on the observation that epidemics typically exploit `heterogeneities' in the population. For networks,, this means that high-degree nodes typically become infected early on in the epidemic. This is the main reason why a first epidemic wave in a network with very heterogeneous degree infects the highly connected nodes. In a scale-free like network, the number of such nodes is small (e.g., 20\% of the nodes responsible for 80\% of infection).  If the epidemic progresses with strong interventions in place, it cannot spread far beyond these high-degree nodes.  Once it dies out, the residual network is highly fragmented and made up of much lower degree nodes.

In this household model, community links drive degree heterogeneity. Household links alone lead to a regular network. Hence, when we effectively cut most community links, heterogeneity in degree is significantly reduced for the duration of control. This means that many high-degree nodes that would normally be infected during the first wave will now not get infected.  The infection is unable to target the highest-degree nodes.  When control is lifted, the high-degree nodes reactivate their links, allowing the epidemic to rebound. 

However, when controlling both link-types equally, degree heterogeneity is preserved and the infection again preferentially targets the high-degree nodes.  A weaker control targeted to all edges rather than just community edges may allow an initial wave of similar size, but it will preferentially target the high-degree nodes.  So this type of control acts as a very effective way of finding highly-connected nodes. This then means that at the end of  lockdown, it is more likely that the most `dangerous' nodes be removed, and with that, a smaller chance of a second wave.


Consider two epidemics with the same intervention start time, providing the optimal strategy for either the global or the community case.  We would expect that the average community degree of those who have been infected in the global case would be higher than in the community case.  Thus, on average, the individuals immunised by infection in the community case will be less important to disease transmission and thus more of them must be immunised to achieve the DIHI threshold.
This is important because households may be able to sustain the epidemic for extended amounts of time and therefore change the outcome in DIHI levels.
\section{Scaling versus modulating the mixing matrix model}\label{sec:results_SEIRD}


In this section, we further explore the notion that modulating the effective $R_0$ of the epidemic through modifying the structure of the mixing matrix (in this case, the age-structure mixing matrix) can affect the system differently from achieving it through simply scaling each element of the mixing matrix. To do so, we began by considering three scenarios described by ~\cite{premProjectingSocialContact2017}, namely, \textit{school closure}, \textit{school closure} and \textit{social distancing}, and \textit{work distancing}. In a baseline, no-intervention case, the matrix of daily contacts $\mathbf{C}$ was set to be the sum of 4 components: \textit{school} contacts, \textit{work} contacts, \textit{home} contacts and \textit{other} contacts. In what follows, we use the corresponding matrices of age-banded daily contacts in the UK produced by the POLYMOD study~\cite{premProjectingSocialContact2017}. Briefly, \textit{school closure} is realised by zeroing the school component of the mixing matrix); \textit{school closure} and \textit{social distancing} involved zeroing school contacts as well as reducing by half the number contacts at other locations between school-going individuals (first four age groups); \textit{work distancing} is implemented by halving the contacts made at the workplace. For each of these interventions, we compared the behaviour of the system when re-scaling the matrix of total contacts so that its $R_0$ during the intervention was the same as the $R_0$ of the modified mixing matrix during the intervention (namely, $2.114$, $2.106$ and $2.179$ for scenarios $1$ to $3$ respectively). 

Age-banded population counts ($18$ $5$-year bands) were taken from the Office for National Statistics (licensed under the Open Government Licence), pooling all age groups above 85. Mortality rates were taken from the modelling of \cite{Verity2020}, assuming that the rates in ten year age bands are the same as across two five year age bands. These rates were calculated based on cases in China in the initial outbreak and from the closed population on the Diamond Princess cruise ship. It is possible that these rates may prove to be overestimates compared to populations that do not experience overwhelming levels of hospitalisation, but they are not likely to be impacted by the interventions considered here. 

To determine the evolution of the epidemic in this baseline case, we first scaled the contact matrices so that the system's $R_0$ was 2.5 (to maintain consistency with previous sections). In all cases, we used the same start date ($T=260$) and duration ($150$ days) for the intervention. These parameters were arbitrarily chosen among the sets of possible parameters resulting in a sub-critical epidemic post-intervention.  





\begin{figure}[] 
    \centering
    \includegraphics[scale=1]{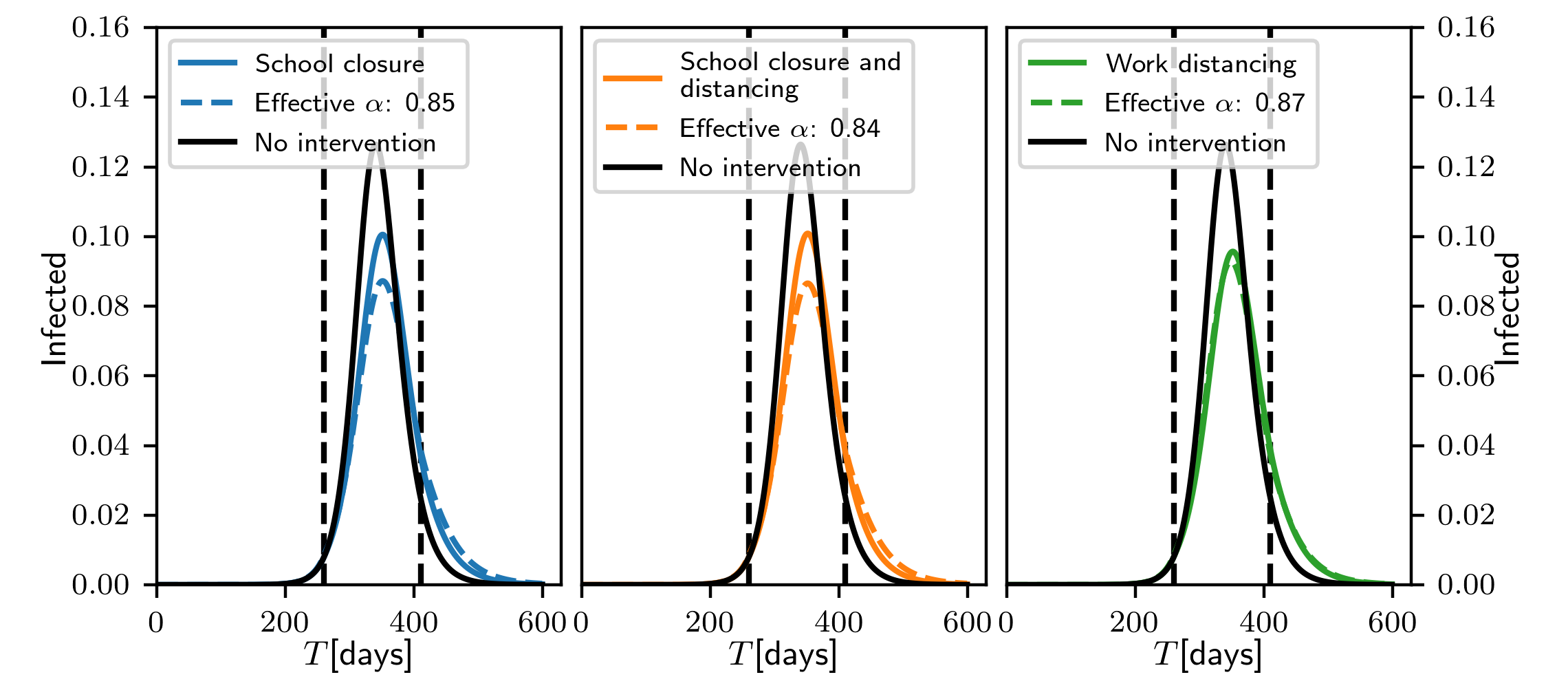}
    \caption{Comparison of the effect of three different control measures in the age-structured compartmental model. The three measures, \textit{school closure}, \textit{school closure} and \textit{social distancing}, \textit{work distancing} (coloured continuous lines, from left to right), act on the structure of the matrix (see text). For reference, the dashed lines result from an intervention reducing infectivity but yielding the same effective $R_0$ during the intervention. Epidemic values are $(\gamma_E,\gamma_I)=(1/7, 1/14)$. Vertical dashed lines indicate beginning and end of control.}
    \label{fig:MD_closures}
\end{figure}

Figure~\ref{fig:MD_closures} confirms that all interventions (coloured lines) result in a reduction in the number of infected individuals. However, whether this intervention is realised through modulating the contact matrix (solid lines) or through uniform scaling (dashed lines) results in substantially different outcomes ($\approx 16\%$ differences for scenarios $1$ and $2$, $\approx 3\%$ in scenario $3$), even though the scaling factors used in the control interventions (dashed lines) are very similar from one scenario to the other. This is clear evidence that the structure of the contacts modulates the effects of the intervention. 

\begin{figure}[]
    \centering
    \includegraphics[scale=1]{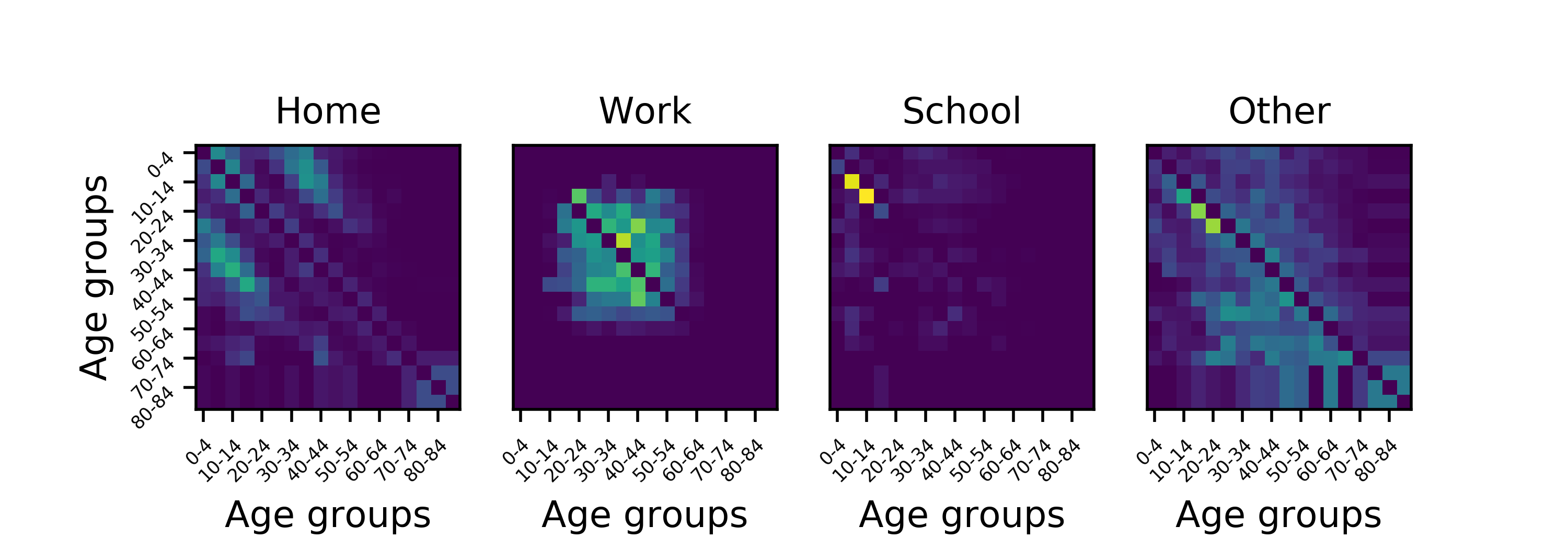}
    \caption{The four components of the POLYMOD age mixing matrix (after subtraction of the diagonal).}
    \label{fig:POLYMOD_components}
\end{figure}

Figure~\ref{fig:POLYMOD_components} provides a visual intuition as to why the \textit{work distancing} scenario is closest to simply scaling the matrix. As pointed out by~\cite{mossongSocialContactsMixing2008}, assortative mixing dominates in $3$ of the components (home, school, other). Thus, \textit{school closure} primarily affects diagonal elements of the mixing matrix (and primarily for the first $4$ age groups). Intervention $2$ does involve halving (some of) the contacts in the \textit{other} component and some of those terms diffuse away from the diagonal, however, these contribute little to the overall mixing matrix. In contrast, the \textit{work} component is the only component to feature what Mossong et al.~\cite{mossongSocialContactsMixing2008} describe as a wide contact plateau. Because this plateau accounts for more than half of the total number of contacts within the corresponding $8$ age-bands, intervention $3$ (social distancing) is most akin to scaling the entire matrix. 

To further clarify how the structure of the contacts modulates the effects of the intervention, we carried out simulations in which two confounding factors were removed, namely heterogeneity in the number of individuals in the different age-bands and in the frequency of contacts by age. Whereas the former plays a key role in the calculation of the effective $R_0$ (see Section~\ref{sec:R0}), the latter weights the impact of the intervention. For example, zeroing \textit{school} contacts which only account for $12\%$ of the total number of contacts will be negligible compared to zeroing \textit{other} contacts that account for $40\%$. Therefore, in what follows, all age groups were set to have the same number of individuals ($1/18$-th of the total population) and all contact components were scaled to have the same sum of elements (arbitrarily, the sum of the original number of \textit{other} contacts). We then systematically analysed the effect of  three different interventions in which one component (\textit{school}, \textit{work}, or \textit{other} was systematically scaled down by a factor taking values between $1.0$ (no intervention) and $0.0$ in steps of $0.1$. 

\begin{figure}[]
\centering
\includegraphics[scale=1]{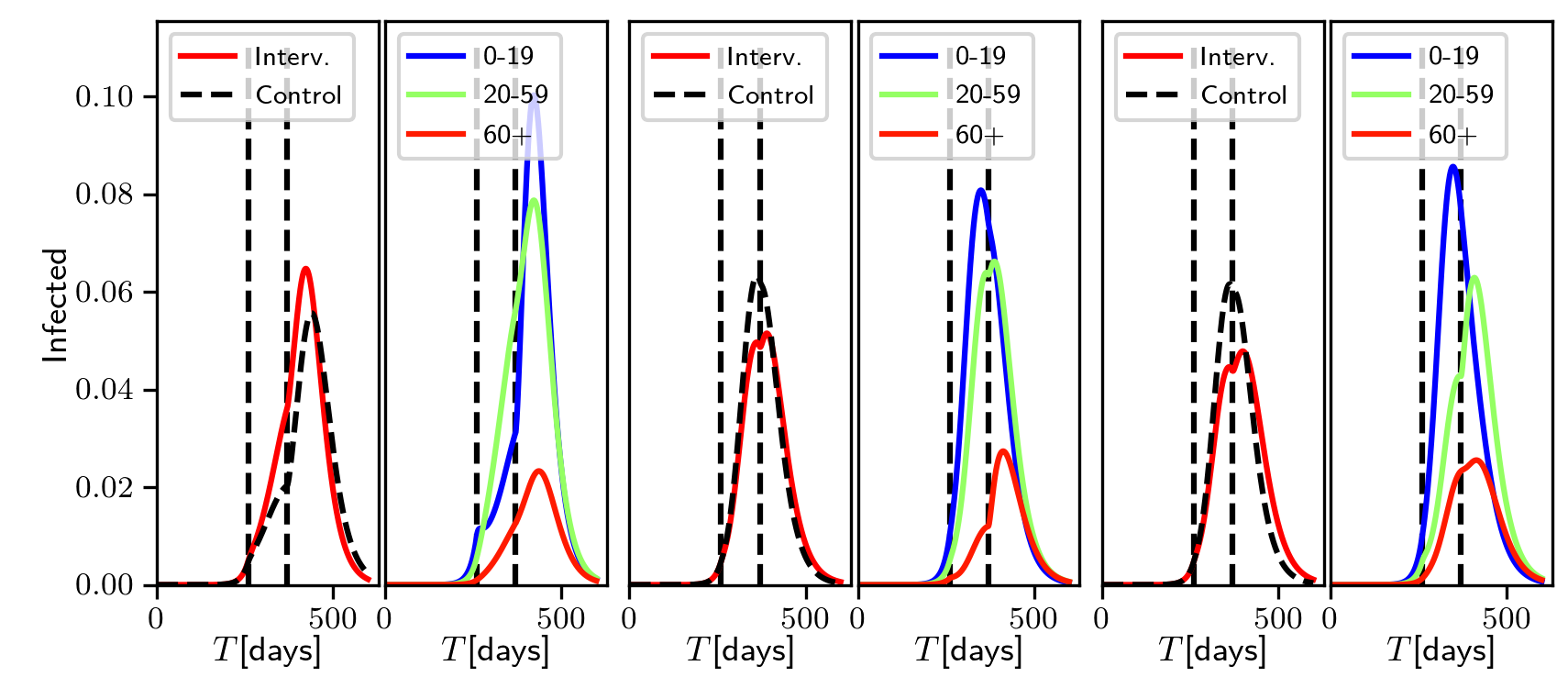}
\caption{Impact of zeroing \textit{school} (left), \textit{work} (middle) and \textit{other} (right) components when each age group has the same number of individuals and each component contributes the same number of contacts. For each, left panel shows the total prevalence of infected individuals in the population using the intervention (red) and the control (scaling of the entire contact matrix to achieve the same effective $R_0$, in black). The right panel shows the prevalence of infected in three pooled age groups chosen to reflect the target of each intervention.}
\label{fig:norm_interv}
\end{figure}

Figure~\ref{fig:norm_interv} shows the effect of the most severe form of intervention (the zeroing of the relevant component) and clearly demonstrates that, once confounding factors are removed, changes in the mixing matrix lead to different outcomes in terms of whether such an intervention is more or less effective than simply scaling the matrix to achieve the same effective $R_0$. Here, zeroing the \textit{school} component is less effective than scaling the overall matrix (despite an overall scaling factor of $0.61$). Instead, zeroing either the \textit{work} or \textit{other} component is more effective than scaling the overall matrix (despite larger scaling factors, $0.87$ and $0.88$, respectively). It should be noted that whilst zeroing the \textit{work} and \textit{other} components leads to similar results in terms of total prevalence, there are differences in prevalence by age which could have significant implications when age-structured mortality rates are considered. 

\section{Discussion}\label{sec:discussion}

In this paper we explored a range of mean-field models previously used to approximate exact epidemics on networks and providing some analytical traction regarding how network properties impact epidemic invasion, final size and the efficacy of control measures. In order of increasing complexity, these models are: the degree-based heterogeneous mean-field, heterogeneous pairwise (without and with clustering), and an edge-based compartmental model which explicitly includes household structure and can distinguish between household and community transmission. While these cannot be used as they are to inform policy they still provide important insights into model selection and key features that need to be captured, or can be exploited, to identify the best possible control measures. In addition, we also tested our findings against a more realistic age-structured model with real mixing matrices.

We have shown that increased degree heterogeneity (i.e., higher variance in the degree distribution) leads to DIHI levels that are much smaller than the basic compartmental model, $1-1/\mathcal{R}_0$. This is line with the findings of~\cite{Britton2020,gomes2020individual}. Moreover, we quantified the extent to which the DIHI induced by the first wave depends on the variance in the degree distribution. We have shown that herd immunity in clustered networks is even lower because epidemics on clustered networks last longer and have lower peaks, allowing more flexibility regarding the start and intensity of control.

Perhaps, the most important question that we addressed regards how lockdown/control is implemented in different models. Many models assume that during lockdown the contact network or mixing matrix is not changing but rather the transmission rate is scaled~\cite{Britton2020,di2020timing,gomes2020individual,morris2020optimal}. We do not believe that this is appropriate because  during lockdown the underlying contact structure changes. Our results with the edge-based compartmental and age-structured models have shown that these two approaches differ in outcome. Perhaps, the assumption of a non-changing contact structure during lockdown is more likely to be made in mean-field models. In models at higher than mean-field resolution (e.g. agent-based) it is much easier to explicitly modify the contact network.

We have therefore revealed a possible risk with using a model that ignores household structure to infer the level of infection required to reach the DIHI threshold.  The favourable change in the DIHI threshold compared to what we predict from a homogeneous model is a consequence of heterogeneities.  Where the intervention makes the population more homogeneous, the disease will no longer act like an effective intervention.  In an extreme case, it will require more infections to achieve herd immunity than a random vaccination would need~\cite{ferrari2006network}. As the infection spreads along edges, we would see that at the end of the first wave, susceptible people are disproportionately in contact with other susceptible people and recovered people are disproportionately in contact with recovered people.  So the residual network of susceptible nodes has more contacts than would occur if the same fraction were vaccinated randomly.

Another important observation resulting from our work is that it is extremely difficult to make general statements by extrapolating from findings based on simple models. Most models in fact ignore meso-scale structures (e.g., degree heterogeneity does well for local or micro structure whilst mixing matrices do well for macro-scale mixing) and their absence may exacerbate the impact of an intervention (either positively or negatively) leading to erroneous conclusions. In the present paper we saw that when intervention could not act on the global network of contacts, DIHI levels varied substantially, although heterogeneities still played a major role in reducing them.\\

Finally, it is worth noting that in all models we considered, when lockdown ended (and if DIHI was not reached), epidemics went on to grow exponentially.  However, in many real-life scenarios, a prolonged phase of sub-critical spreading (i.e. slow decay) has been observed before the exponential growing phase returned. There is a number of reasons why this behaviour is not observed in our models: (i) Deterministic  models fail to capture fluctuations that dominate when the number of infected people is small, which might have a major impact on resurgence; (ii) during lockdown, the contact structure changes drastically and abruptly, but when lockdown is lifted, there is a delay/inertia in going back to pre-lockdown status, such lag having important implications on resurgence; (iii) after lockdown, some social distancing measures remain along with social awareness reducing exposure to the disease. Understanding and modelling these effects is an interesting challenge to address in future work.

In complementing our study of network-based mean-field models, we used a model with no explicit contact structure, where instead contact structure was alluded to via age-related mixing patterns. In the present situation with Covid-19, such models are appealing because they explain some of the structure of the population and provide a rationale for establishing a lockdown. Moreover, with higher mortality rates among older people, age-structured models are of interest in their own right. There is some tension between which model is most apt for describing population experience of the infection, the control measures effected and the outcomes for the population. The ideal solution may lie between some of the options presented here.

However, age-structured models cannot say much that is explicit about the structure of contacts within the population. While we have discussed ways in which lockdowns can be implemented in such models, the formulation is arguably less intuitive than in the network case. How to bridge the gap is an interesting question. An age-related network might be one labelled with age classes, with analysis focused on understanding the positions in the network occupied by individuals of a given age. This could be coupled to household models of network formation, one with variable sizes and smaller households more likely to feature older individuals. Once the general structure of such networks is known, adapting the models here would be simple enough. These considerations are beyond the presentation here but would be a fruitful avenue for future discussions, particular if additional lockdowns are required.

\newpage

\section{Appendix}
\subsection{Pairwise equations}\label{sec:het-PW-eq}

We can account for clustering in the pairwise model, by introducing a clustering factor $\varphi$. The equations are similar to~\eqref{eq:pairwise}, but when closures are implemented we have to consider both open and closed triples.\\

In the case of open triples, the equations for the closures are the same as in \eqref{eq:pairwise}, scaled by $(1-\varphi)$ to account for clustering, i.e. 
\begin{eqnarray}
&[A_\ell S_kI]^o =(1-\varphi) \frac{k-1}{k} \frac{[A_\ell S_k][S_kI]}{[S_k]},\nonumber\\
&[IS_kA_\ell ]^o =(1-\varphi) \frac{k-1}{k} \frac{[I S_k][S_kA_\ell ]}{[S_k]},\nonumber\\
\label{eq:opentriplesclosures}
\end{eqnarray}
where the superscript $^o$ indicates open triples.
For closed triples, we first study $[A_\ell S_kI_m]^c$. This quantity represents triples in which the first node is in state $A$, the middle node is in state $S$ with degree $k$, and the third node is in state $I$ with degree $m$. The triple is closed, therefore the third node is connected to the first one. This introduces correlations when we write the triple in terms of pairs for the closure, which we indicate with $C_{S_\ell I_M}$.

\begin{eqnarray}
[A_kS_\ell I_m] = \varphi [A_kS_\ell ](\ell -1) \frac{[S_\ell I_m]}{\ell [S_\ell ]} C_{A_kI_m}.
\end{eqnarray}
The correlation can be written in terms of the ratio between realized pairs $[A_kI_m]$ and possible pairs $[A_kI_m]$ in a well-mixed population:
\begin{equation}
    C_{A_kI_m} = \frac{[A_kI_m]}{k  [A_k] \frac{m[I_m]}{\langle k \rangle N}}.
\end{equation}
Hence, the closure is
\begin{eqnarray}
[A_kS_\ell I_m] = \varphi \frac{(\ell -1)}{\ell }\frac{\langle k \rangle N}{km}\frac{[A_k S_\ell ][S_\ell I_m][A_k I_m]}{[A_k][S_\ell ][I_m]}.
\end{eqnarray}
In a similar manner, we can write the expression for $[I_mS_kA_\ell ]$. The resulting system is therefore
\begin{eqnarray}
 \dot{[S_k]} &=& -\tau [S_kI] \nonumber \\
 \dot{[I_k]} &=& \tau [S_kI] - \gamma[I_k]\nonumber \\
 \dot{[R_k]} &=& \gamma[I_k]\nonumber \\
 \dot{[S_kI_\ell ]} &=& - \gamma[S_kI_\ell ] + \tau \left[ \sum_m \left( [S_kS_\ell I_m] -[I_mS_kI_\ell ] \right) -[S_kI_\ell ] \right] \nonumber \\
 \dot{[S_kS_\ell ]} &=& - \tau\left[ \sum_m \bigg( [S_kS_\ell I_m] -[I_mS_kI_\ell ] \bigg) \right],\nonumber \\
 \label{eq:clustered}
\end{eqnarray}
where
\begin{equation*}
[A_kS_\ell I_m] = \varphi [A_kS_\ell ]\frac{\ell -1}{\ell } \frac{[S_\ell I_m]^2}{ \langle k \rangle [S_\ell ]^2 \frac{[I_m]}{N}} + (1-\varphi)  \frac{\ell -1}{\ell } \frac{[A_kS_\ell ][S_\ell I_m]}{[S_\ell ]}. \nonumber
\end{equation*}

\subsection{Edge-based compartmental model}\label{sec:ebcm-eq}
We consider a network with $4N$ nodes partitioned into households of size $4$. Apart from the within household community, each node has a number of links to nodes outside the household, according to the degree distribution $P_{n,p}(k)$. The within household per-contact-transmission is denoted by $\beta_h$ while the community transmission by $\beta_c$. The resulting system is given below,

\begin{eqnarray}
 &\dot{\varphi}_{SSS}(t)& = - 3A(t)\varphi_{SSS}(t),\\
 &\dot{\varphi}_{SSI}(t)& = 3A(t)\varphi_{SSS}(t) - \gamma \varphi_{SSI}(t) - 2(A(t)+\beta_h) \varphi_{SSI}(t) - \beta_h \varphi_{SSI}(t),\\
 &\dot{\varphi}_{SSR}(t)& = \gamma \varphi_{SSI}(t) - 2A(t) \varphi_{SSR}(t),\\
 &\dot{\varphi}_{SIR}(t)& = 2A(t) \varphi_{SSR}(t) + 2\gamma \varphi_{SII}(t) - \gamma \varphi_{SIR}(t) - \beta_h \varphi_{SIR}(t),\\
 &\dot{\varphi}_{SRR}(t)& = \gamma \varphi_{SIR}(t) - A(t) \varphi_{SRR}(t),\\
 &\dot{\varphi}_{IRR}(t)& = A(t) \varphi_{SRR}(t) + 2 \gamma \varphi_{IIR}(t) - \gamma \varphi_{IRR}(t) - \beta_h \varphi_{IRR}(t),\\
 &\dot{\varphi}_{RRR}(t)& = \gamma \varphi_{IRR}(t),\\
 &\dot{\varphi}_{IIR}(t)& = -2 \gamma \varphi_{IIR}(t) - 2 \beta_h \varphi_{IIR}(t) + 3\gamma \varphi_{III}(t),\\
 &\dot{\varphi}_{III}(t)& = (A(t)+2\beta_h) \varphi_{SII}(t) - 3 \gamma \varphi_{III}(t) - 3\beta_h \varphi_{III}(t),\\
 &\dot{\varphi}_{SII}(t)& =-(A(t)+2\beta_h) \varphi_{SII}(t) - 2\beta_h\varphi_{SII}(t) + 2(A(t)+\beta_h) \varphi_{SSI}(t) - 2\gamma \varphi_{SII}(t),\\
 &\dot{\Theta}(t)& = -(  \beta_h \varphi_{SSI}(t) + 2\beta_h \varphi_{SII}(t) + 3\beta_h \varphi_{III}(t) + \beta_h \varphi_{SIR}(t) + 2\beta_h \varphi_{IIR}(t) + \nonumber\\
 && \quad +\beta_h \varphi_{IRR}(t)),\\
 &\dot{\theta}(t)& = - \beta_c \varphi_I(t),\\
 &\varphi_I(t)& = \theta(t) - \gamma(1-\theta(t))/\beta_c - (1-\epsilon) \frac{\varphi^\prime (\theta(t))}{\langle k \rangle} \Theta(t),\\
 &A(t)& =  \frac{\psi^\prime (\theta(t))}{\psi(\theta(t))} \beta_c \varphi_I(t),\\
 &S(t)& = (1-\epsilon) \Theta(t) \psi(\theta(t)),\\
 &\dot{R}(t)& = \gamma I(t),\\
 &I(t)& = 1 - S - R,
\end{eqnarray}
with the following initial conditions:
\begin{eqnarray}
 \varphi_{SSS}(0) = (1-\epsilon)^3,\\
 \varphi_{SSI}(0) = 3\epsilon (1-\epsilon)^2,\\
 \varphi_{SII}(0) = 3\epsilon^2 (1-\epsilon),\\
 \varphi_{III}(0) =\epsilon^3,\\
 \theta(0) = 1,\\
 \Theta(0) = 1,\\
 S(0) = 1-\epsilon,\\
 I(0) = \epsilon,\\
\end{eqnarray}
with all other variables set to zero at time $t=0$.

\clearpage

\section{Acknowledgments}
F. Di Lauro, L. Berthouze and I.Z. Kiss acknowledge support from the Leverhulme Trust for the Research Project Grant RPG-2017-370.  J.C Miller acknowledges startup funding from La Trobe University.

\clearpage
\bibliography{biblio}
\bibliographystyle{plain}
\end{document}